\documentclass[twocolumn]{aa}
\usepackage{graphicx,amsmath,amssymb}
\begin{document}
   \title{$Chandra$ Discovery of an X-ray Jet and Lobes in 3C~15}


   \author{J. Kataoka\inst{1},
          J. P. Leahy\inst{2}, 
          P. G. Edwards\inst{3},
          M. Kino\inst{4,5},
          F. Takahara\inst{5}\\
          Y. Serino\inst{1},
          N. Kawai\inst{1}, and 
          A. R. Martel\inst{6}
   }

   \authorrunning{Kataoka et al.}

   \offprints{J. Kataoka}
   \institute{Tokyo Institute of Technology,
              Meguro-ku, Tokyo, Japan\\
              \email{kataoka@hp.phys.titech.ac.jp}
         \and
              University of Manchester, Jodrell Bank Observatory,   
              Macclesfield, Cheshire SK11 9DL, United Kingdom 
         \and
              Institute of Space and Astronautical Science, Sagamihara,
              Kanagawa, Japan
         \and
              Earth and Planetary Science, University of Tokyo 
         \and
              Department of Earth and Space Science, Osaka University, 
              Osaka, Japan
         \and
              Space Telescope Science Institute, 3700 San Martin Drive,
              Baltimore, MD 21218, USA
             }

   \date{Received 6 June 2003 / Accepted 12 August 2003}

   \abstract{
We report the $Chandra$ detection of an X-ray jet in 3C~15.
The peak of the X-ray emission in the jet is 4.1$''$ (a projected
distance of 5.1~kpc) from the nucleus, and coincident with a component
previously identified in the radio and optical jets.
We construct the spectral energy distribution (SED) for this
component, optical knot~C, and find that X-ray flux is well below the
extrapolation of the radio-to-optical continuum.
We examine four models for the X-ray jet emission:
(I) weak synchrotron cooling in equipartition, 
(II) moderate synchrotron cooling in equipartition, 
(III) weak synchrotron plus synchrotron self-Compton (SSC) cooling, and 
(IV) moderate synchrotron plus SSC cooling.  
Given weak evidence for a concave feature in the X-ray spectrum, we
argue that case (II) can most reasonably explain the overall emission
from knot~C.
Case (III) is also possible, but requires a large departure from
equipartition and for the jet power to be comparable to that of the
brightest quasars.
In all models, (I)$-$(IV), electrons must be accelerated 
up to $\gamma_{\rm max} \gtrsim 10^7$, 
suggesting that re-acceleration is necessary in knot~C of the 3C~15 jet. 
Diffuse X-ray emission has also been detected, distributed widely
over the full extent (63~kpc$\times$25~kpc) of the radio lobes. The
X-ray spectrum of the diffuse emission is described by a two-component model,
consisting of soft thermal plasma emission from the host galaxy
halo and a hard nonthermal power-law component. The hard component
can be ascribed to the inverse Comptonization of cosmic microwave
background (CMB) photons by the synchrotron emitting electrons
in the radio lobes. 
We compare the total energy contained in the lobes with the jet power
estimated from knot C, and discuss the energetic link between the jet
and the lobes.  We argue that the fueling time ($t_{\rm fuel}$) and
the source age ($t_{\rm src}$) are comparable for case (II), whereas
$t_{\rm fuel}$ $\ll$ $t_{\rm src}$ is likely for case (III). 
The latter may imply that the jet has a very small filling factor,
$\sim$10$^{-3}$.  We consider the pressure balance between the thermal
galaxy halo and non-thermal relativistic electrons in the radio lobes.
Finally, we show that the X-ray emission from the nucleus is not
adequately fitted by a simple absorbed power-law model, but needs an
additional power-law with heavy absorption ($N_{\rm H} \simeq
10^{22-23}$~cm$^{-2}$) intrinsic to the source.  Such a high column
density is consistent with the presence of a dense,
dusty torus which obscures the quasar nucleus.

   }

   \maketitle
%

\section{Introduction}

Jets are among the most exciting (but also among the least understood) 
cosmic phenomena, being highly efficient particle accelerators that
generate relativistic electron distributions. This extreme jet power is
most likely generated near a super-massive black hole, probably via the
conversion of gravitational energy from accreting matter.
The interest in observations of jets at shorter wavelengths is
related to the ability to (1) probe the sites of high energy particle 
acceleration and help determine their physical parameters, and  
(2) trace the energy transport from the nucleus to the outer hot 
spots/radio lobes.  However, observations of jets with high spatial 
resolution ($\le 1''$) have until recently only been possible at
lower frequencies, i.e., at radio and optical wavelengths.

The excellent spatial resolution of the $Chandra$ $X$-$Ray$ $Observatory$
has now resolved the X-ray spatial structure along the jets of more than 
20 radio galaxies (e.g., Harris and Krawczynski 2002, and references
therein).  The broadband spectral energy distributions (SEDs) of knots
and hotspots show great variety between the radio and X-ray energy
bands.  In most cases, the X-ray spectra are much harder than expected 
from a simple extrapolation of the optical fluxes. This suggests that both 
the radio and optical photons are due to the same non-thermal synchrotron
radiation, whereas X-ray photons are produced via Compton scattering 
of either synchrotron photons (SSC; e.g., Wilson, Young, \& Shopbell
2001 for the hotspot of Pictor~A) and/or cosmic microwave background 
photons (EC/CMB; e.g., Sambruna et al.\ 2001 for the knots in 3C~273). 
In other cases, the X-ray--optical--radio spectrum is consistent with a 
single smoothly broken power-law spectrum, suggesting that the broad 
band emission is entirely due to non-thermal synchrotron radiation 
(e.g., Wilson \& Yang 2002 for the jet of M87, and Hardcastle, Birkinshaw 
\& Worrall 2001 for jet in 3C~66B).

The X-ray detections of jets and hotspots have a great impact on the
determination of their physical parameters, as well as the maximum energy of 
accelerated electrons (see (1) above). Once the source of seed 
photons is identified, a comparison of the synchrotron (radio to X-ray) 
and inverse Compton (SSC or EC/CMB; X-ray) fluxes allows us to
independently determine the energy densities of the relativistic electrons and 
the magnetic field, $u_{\rm e}$ and $u_B$ respectively. This approach has
been applied to X-ray observations of radio lobes with $ASCA$
(Kaneda et al.\ 1995;  Tashiro et al.\ 2001), in which the X-ray emission
extends over arcmin scale sizes. By analyzing X-ray data of more than 
10 radio lobes, Isobe (2002) found that lobes are generally 
``particle dominated'', in the sense that $u_{\rm e} \ge u_B$.   
Recent $Chandra$ detections of knots and hotspots have resulted in this
idea being extended to more compact regions in jets with $\sim$arcsec
spatial extents.  In some objects, the X-ray fluxes of knots and hotspots 
turned out to be much brighter than that expected from equipartition 
between particles and fields (e.g., Hardcastle et al.\ 2002, 
Kataoka et al.\ 2003; Georganopoulos \& Kazanas 2003). 

Comparison of the powers in jets and lobes provides an important 
clue to the formation of astrophysical jets and the evolution of radio 
galaxies (see (2) above). This is because lobes are undoubtedly 
$fueled$ by jets over a long time, possibly throughout their lifetime.
Assuming the total power of the jet ($L_{\rm E}$) and the total 
energy supplied in the radio lobes ($E_{\rm lobe}$),  
the ``fueling time'' of the system may be defined as 
$t_{\rm fuel} \sim E_{\rm lobe}/L_{\rm E}$. 
Meanwhile, the ``source age'' can be defined as 
$t_{\rm src} \sim D/v_{\rm exp}$, where the $D$ is the 
spatial extent and $v_{\rm exp}$ is the expansion speed of the radio
lobes. Observationally, symmetry arguments show that the main-axis
expansion of FR II lobes is sub-relativistic 
($v_{\rm exp}$ $\le$ 0.1 $c$\,; e.g., Scheuer 1995), 
and since the lobes are not spherical the
limit to transverse expansion is several times lower. In fact, lobes are 
surrounded by hot plasma which will prevent them expanding freely, 
either by thermal or by ram-pressure confinement (e.g., Leahy \& Gizani
 2001).
We expect $t_{\rm fuel} \sim t_{\rm src}$ if the jet power is 
injected into the radio lobes in a stationary, constant manner.  
A large departure from this equality may imply that either the injection 
of power by the jet is not uniform, and/or that there are significant 
contributions from particles which we cannot observe. 
Comparison of $t_{\rm fuel}$ and $t_{\rm src}$ 
provides important information for understanding the jet-lobe connection.

3C~15 is an unusual radio source with an optical jet 
which has been imaged with the $Hubble$ $Space$ $Telescope$ ($HST$).
The $HST$ observations found no bright, 
unresolved nucleus in the core of the galaxy,
in contrast to other galaxies hosting optical jets (Martel et al.\ 1998). 
The radio structure of 3C~15 is intermediate between Fanaroff--Riley (FR)
classes I and II, 
although its luminosity is above that of a number of sources 
showing classical FR~II structure (Leahy et al.\ 1997). The optical 
morphology of the jet matches the features of the radio jet very well, 
suggesting that the optical emission is strongly dominated by
synchrotron radiation.
In this paper, we report the $Chandra$ discovery of X-ray emission
from the knots, lobes and nucleus of 3C~15.
By combining the data from radio to X-rays, we derive the power of the 
jet and the lobes separately for extensive study. 
Throughout this paper, we adopt $H_0 = 75$~km\,s$^{-1}$\,Mpc$^{-1}$ and
$q_0 = 0.5$, so that 1$''$ corresponds to 1.25~kpc at the redshift of
3C~15, $z=0.073$.

\section{Observations}
\subsection{Summary of Previous Radio Observations}

3C~15 has a bright northern jet containing four prominent knots,
with a much weaker counter-jet on the southern side of the nucleus.
VLA observations at 8.3\,GHz were made by Leahy et al.\ (1997), and 
we use these data in this paper, with an 0.34~arcsec resolution image shown
as contour levels in Figure~1 ($upper$ $right$). 
The nucleus and jets are surrounded by lobe emission, which extends over 
50~arcsec (a projected distance of 63~kpc). The surface brightness profile
of 3C~15 projected onto the major axis of the radio lobes
is shown in Figure~2. Leahy et al.\ (1997) labeled the four radio knots 
A, B, C and D, however, the optical observations described in the next 
section resolved the radio knot~A into two optical knots. In this paper 
we follow the optical scheme of labeling the knots A and B (corresponding
to radio knot~A) and C (corresponding to radio knot~B).
The jet--counter-jet flux density ratio is 50$\pm$10 (Leahy et al.\ 1997).
The VLA images suggests the source does not lie in the plane of the sky,
in which case the jet asymmetry may partially be due to relativistic
beaming. Assuming a spectral index, $\alpha_{\rm R}$, of 0.6 (where $S \propto 
\nu^{-\alpha_{\rm R}}$), the flux asymmetry implies $\beta \cos \theta = 
0.635 \pm 0.025$,  hence $\theta \simeq 45$--$50^{\circ}$ for $\beta \ge 0.9$. 
This corresponds to a Doppler beaming factor in the range
$0.40 \le \delta \le 1.2$. 

   \begin{figure*}
   \centering
   \includegraphics[width=7.4cm]{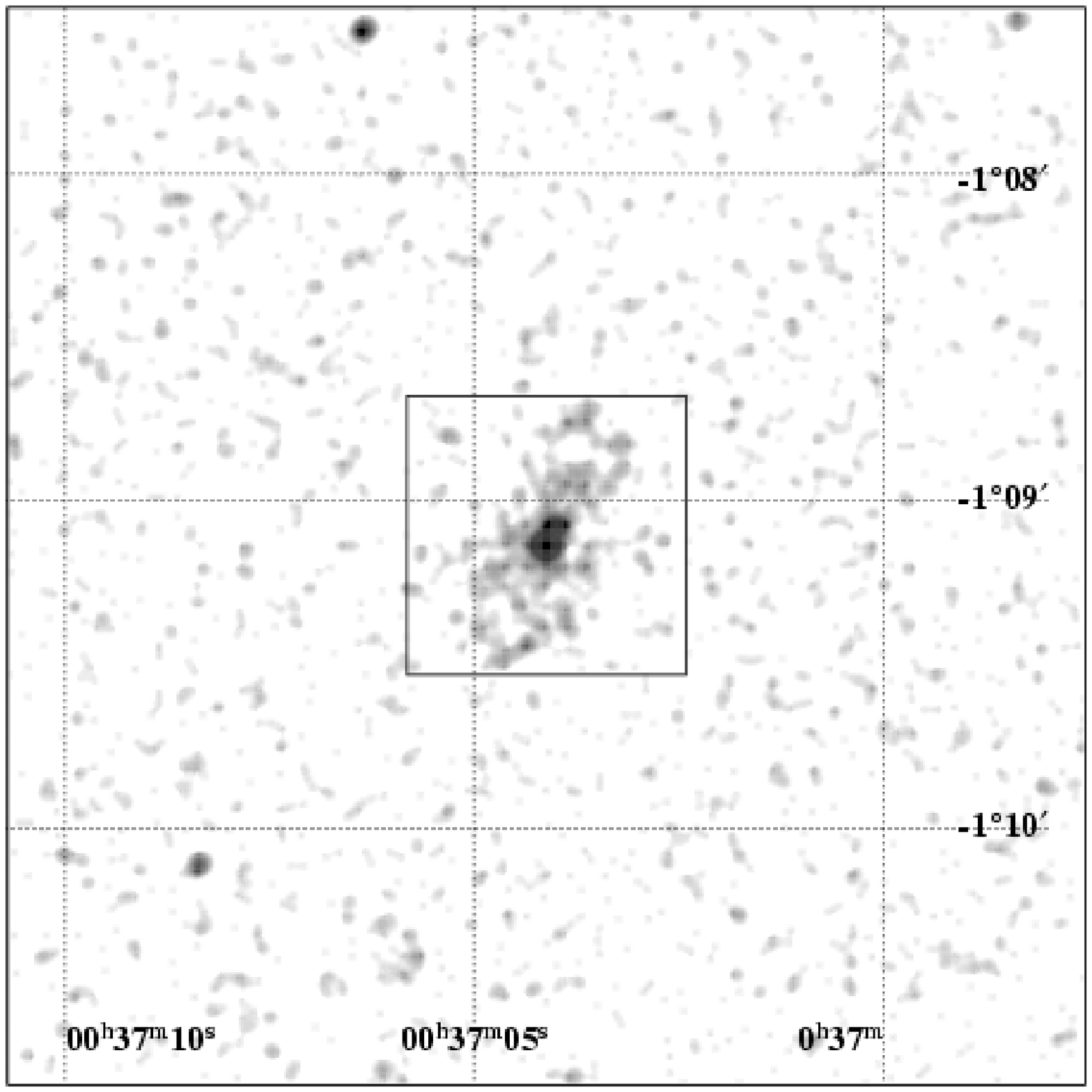}
   \includegraphics[width=7.4cm]{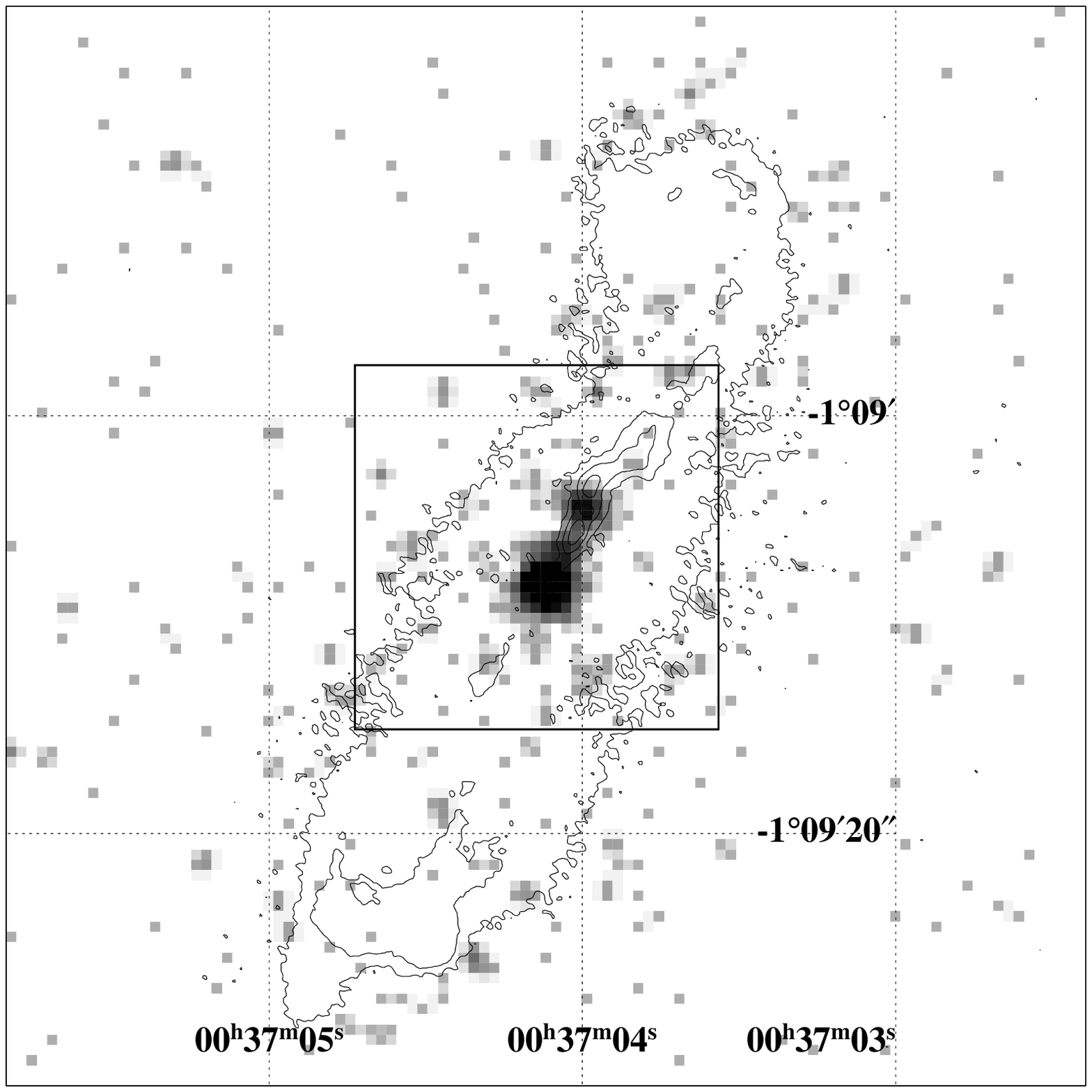}
   \includegraphics[width=7.4cm]{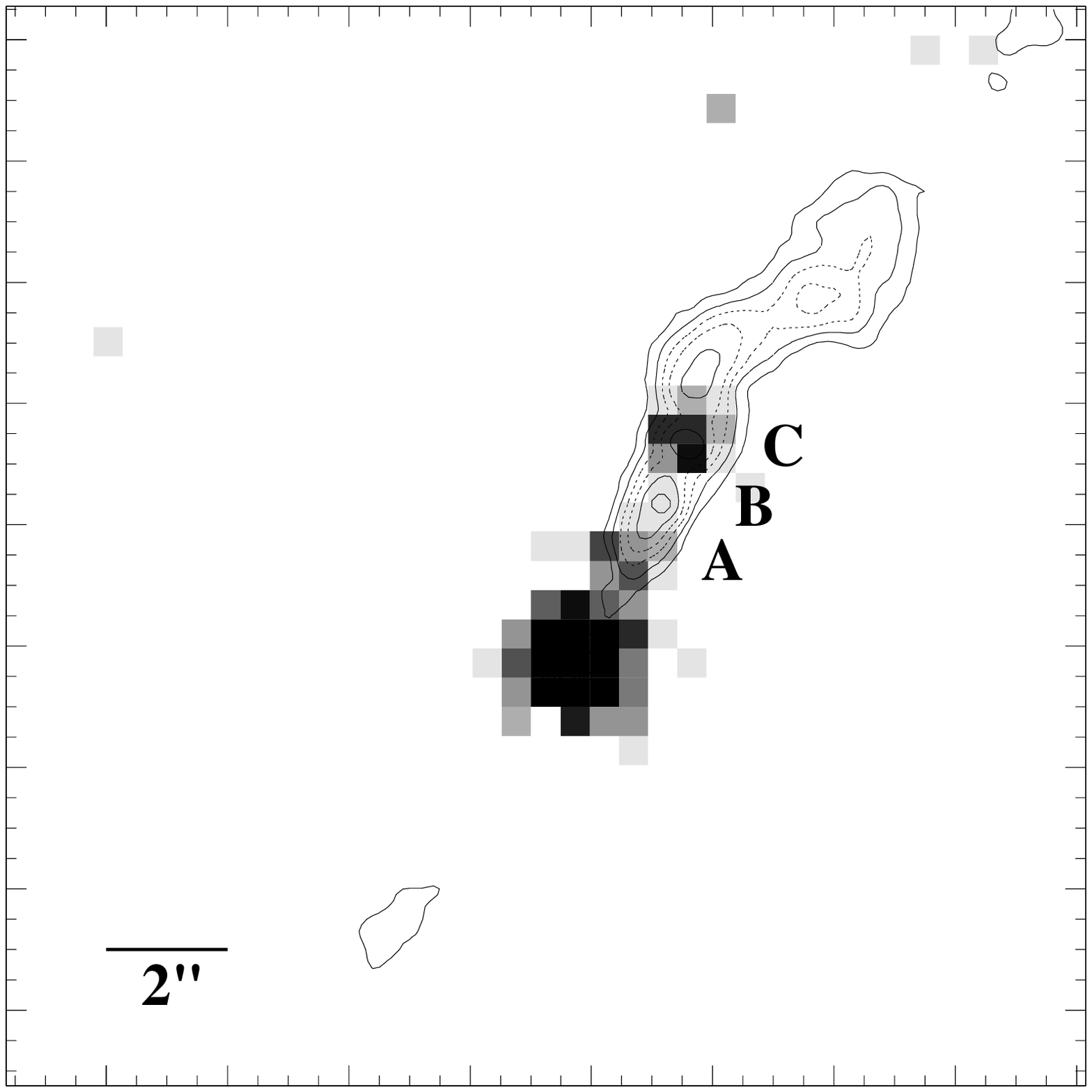}
      \caption{$Upper$ $left$: A 0.4--8~keV X-ray image of 3C~15 
($Chandra$ ACIS-S3) smoothed with a two-dimensional Gaussian of 
$\sigma = 1.0$~arcsec. $Upper$ $right$: Radio and X-ray image of 3C~15 
lobes and jet (expanded plot of the solid square in the $left$ panel). 
The grey scale is the X-ray image smoothed with a 
$\sigma = 0.3$~arcsec Gaussian. The VLA 8.3\,GHz image is adapted from 
Leahy et al.\ (1997). The contour levels are (0.02, 0.11, 0.62, 3.5, 20) 
mJy beam$^{-1}$ for a beam size of 0.34$'' \times 0.34''$.  
$bottom$: Expanded plot of the central jet region (17.4$'' \times 17.4''$) 
shown by the solid square in the $upper$ $right$ panel).
The image is smoothed with a $\sigma = 0.3$~arcsec Gaussian. A, B, and C
denote the jet knots as defined in Martel et al.\ (1998).
The VLA contour levels are (0.2, 0.56, 1.6, 4.3, 12) mJy beam$^{-1}$ for a 
beam size of 0.34$'' \times 0.34''$. } 
         \label{BGreg}
   \end{figure*}

\subsection{Summary of Previous Optical Observations}

The optical counterpart of the radio jet in 3C~15 was found in $HST$ 
observations by Martel et al.\ (1998), and we use these $HST$ data in 
this paper. The optical morphology of the jet closely matches the 
features of the radio jet, suggesting that the optical emission is 
strongly dominated by synchrotron radiation (Figure~5 of Martel et al.\ 1998).
Three prominent knots in the jet were detected at a P.A.\ of
$-$30$^{\circ}$.  Thanks to the excellent image resolution of the 
$HST$ (0.1$''$), the size of each knot has been accurately measured. 
The innermost knot, knot~A, is broad and diffuse. It extends from 
2.6~kpc (2.1$''$) to 3.5~kpc (2.8$''$) from the nucleus. Knots B and C 
are comparatively compact and well defined. The FWHMs of knot B and C
are both $\sim$0.5~kpc. The outermost knots of the radio jet, defined as
C and D by Leahy et al.\ (1997) are not detected in the optical image.  
Martel et al.\ (1998) compared the fluxes of the jet measured in the 
$V$ and $R$ bands with radio and soft X-ray data. The low resolution
radio data and the X-ray observations made by $ROSAT$ include flux 
contributions from both the jet and the nucleus of 3C~15. A fit through 
the radio to optical data yields $\alpha_{\rm ro} = 1.03\pm0.02$. 
When the flux measurement of only the northern jet (combined knots A, B, 
and C) is considered, a radio-optical spectral index of 0.95$\pm$0.01 
is obtained.

\subsection{X-ray Observation and Data Reduction}

3C~15 was observed by the $Chandra$ $X$-$ray$ $Observatory$ on 2000 November 6 
in a guaranteed time observation (sequence number 700368, obs.\ ID
2178) using the Advanced CCD Imaging Spectrometer (ACIS-S) spectroscopic 
array.  The nucleus of 3C~15 was centered 20$''$ in the $-$Y direction
from the location of the best focus on chip S3 (nominal position of the 
ACIS-S3 chip).
All of the regions of radio emission from 3C~15 were imaged on S3.
The total good time interval was 28.2~ksec, taken with the default frame
time of 3.2 sec. 
We have analyzed archival data on 3C~15 provided by HEASARC Browse 
(http://heasarc.gsfc.nasa.gov/dp-perl/W3Browse/Browse.pl).
The raw level-1 data were reprocessed using the latest version
(CIAO~2.2.1) of the CXCDS software and version 11.1.0 of XSPEC.  
We generated a clean data set by selecting the standard grades 
(0, 2, 3, 4 and 6) and the energy band 0.4$-$8~keV.

\section{Results}
\subsection{X-ray Image}

The X-ray image, produced by smoothing the raw $Chandra$ image 
in the energy range of 0.4$-$8 keV with a Gaussian of width 1.0$''$ 
is shown in Figure~1 ($upper$ $left$). We find faint X-ray emission 
surrounding the nucleus, extended in the north-west--south-east
direction. Comparing this to the 8.3\,GHz VLA image in Leahy et al.\ (1997), 
the extended X-ray emission appears to be correlated with the extended
radio lobes. An expanded plot of the solid square in the left figure is shown in 
the right panel, where the radio contours are superposed on the 
X-ray image (grey scale) smoothed with a $\sigma$ = 0.3 arcsec Gaussian.  
X-ray emission can be clearly seen from the bright central nucleus and 
the northern jet. An expanded plot of the central jet region 
(17.4$''\times$17.4$''$) is shown in the bottom panel. 
The nucleus and jet knots are detected at the 22.9\,$\sigma$ (527 counts) 
and 7.7\,$\sigma$ (60 counts for knot~C: see below) level, respectively. 
The number of net X-ray photons associated with the radio lobes is 246
photons, which includes a background of 73 photons in the 0.4$-$8 keV
range (11\,$\sigma$ level), where we have excluded pixels within 1.9$''$ 
of the possible 
point sources, including knot~C, knot~A$'$ and the nucleus (see below). 

   \begin{figure}
   \centering
   \includegraphics[width=7.0cm,angle=90]{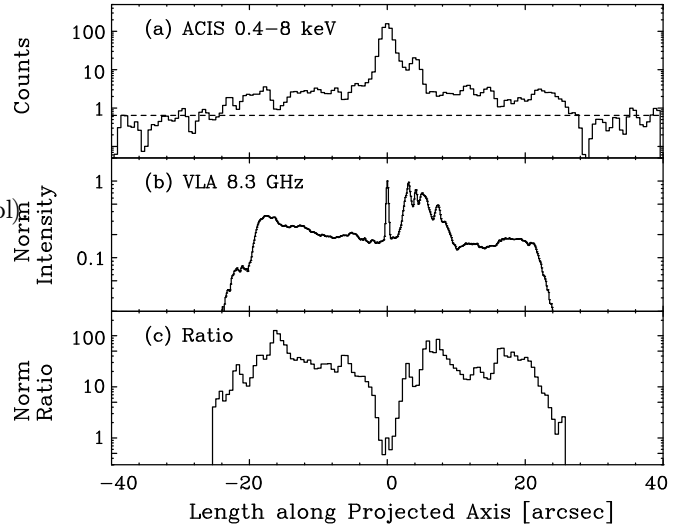}
      \caption{Log profiles along the major axis of the lobes of 3C~15.
    (a) the X-rays in the 0.4$-$8~keV range, (b) the radio emission at
    8.3~GHz (normalized to the nuclear radio flux), 
    and (c) the ratio between the radio and X-ray profiles 
($S_{\rm radio}$/$S_{\rm X}$; normalized to central nucleus).}
         \label{SED}
   \end{figure}

   \begin{figure}
   \centering
   \includegraphics[width=7.0cm,angle=90]{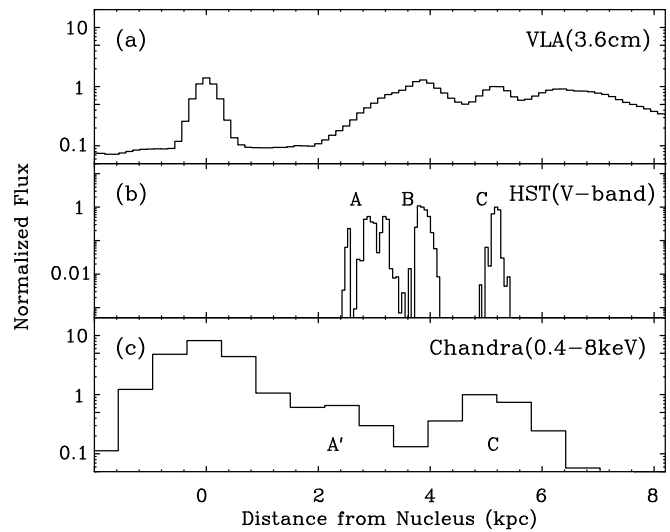}
      \caption{Log profiles along the major axis of the jet of 3C~15,
(a) the radio emission at 8.3~GHz,
(b) the $V$ band ($HST$; adapted from Fig 3 in Martel et al. 1998), and 
(c) the X-rays in the 0.4$-$8 keV range.
All jet profiles are normalized to the knot~C flux. 
The underlying galactic light, and hence bright nuclear
emission, are subtracted in the optical profile (b).}
         \label{SED}
   \end{figure}

In order to examine the spatial correlation between the radio and  X-ray 
emission, we projected the raw ACIS image onto the major axis of the 
radio lobes. Here the strip integration is restricted to a width of 
20$''$ transverse to the major axis.
 The X-ray logarithmic surface brightness profile 
is shown in Figure~2(a), together with the radio profile 
in Figure~2(b).  The X-ray profile has been smoothed with a 
Gaussian of width 1$''$. No background subtraction has been performed in 
the X-ray image, hence the profile contains an average background of 
0.64~counts/bin in the 0.4$-$8.0 keV band (given as the dashed line in 
Figure~2(a)).
Although the X-ray photon statistics are limited, the X-ray and radio profiles 
clearly show similar spatial extents but different structures. 
The X-ray profile exhibits a ``center-filled'' morphology with a 
relatively uniform surface brightness (except for the nucleus region), 
while the northern jet and southern rim are bright in the radio image. 
These differences can also be seen in the bottom panel, where the 
surface brightness ratio
 ($S_{\rm radio}$/$S_{\rm X}$; normalized to the nucleus) 
is a minimum at the center, and gradually increases at larger distances.  

The expanded profile of the X-ray jet is shown in Figure~3, together 
with the radio (VLA 8.3~GHz) and optical ($HST$) jet profiles. 
Note that the strong background light of host galaxy is subtracted in the
optical profile, resulting in only the jet knots being visible
(see also \S4.3).
The X-ray image resolution is poorer than that of the optical and radio, but 
it is clear that most of the X-ray jet emission originates 
from the knot~C of Martel et al.\ (1998). The total extent (FWHM) of the 
X-ray nucleus and knot~C are both $\sim$1$''$, 
consistent with the the broadening 
of the $Chandra$ PSF ($\sim 0.5''$ half-energy radius). 
There is also a suggestion of faint X-ray emission 
from the northern jet intermediate between the nucleus and knot~C
(denoted A$'$ in Figure 3 (c)), although this is heavily  
contaminated by the bright nuclear emission. 
This feature is located 1.8$\pm$0.1$''$ from 
the nucleus, and contains 45$\pm$12 photons. 
This position is
approximately coincident with the innermost knot detected 
by $HST$ (knot~A; see $\S$2.2), however the offset of (0.2$-$0.9)$\pm$0.1$''$ 
between the X-ray and optical peaks, with the X-ray peak being closer 
to the nucleus, seems to be significant. 
Future observations with improved photon statistics and image resolution
are necessary to confirm this.

\subsection{X-ray Spectrum of knot C}

We extracted the X-ray photons from knot C 
within a circular region of 1.6$''$ radius, to reduce the contamination 
of photons from knot~A$'$ (to less than $\sim$10\,\%).
Approximately 97$\%$ of the counts from a point source are 
collected within a circular region of this size. 
Background subtraction was performed, where the
background counts were accumulated at the same off-nuclear distance. 
Figure~4 shows the background-subtracted ACIS
spectrum of knot~C. We first assumed a power-law function absorbed by 
Galactic $N_{\rm H}$ only (fixed to 3.03$\times$10$^{20}$~cm$^2$: Stark
et al.\ 1992). The best fit X-ray photon index was
 $\Gamma_{\rm jet}$ = 1.71$^{+0.44}_{-0.36}$, 
and the corresponding 0.5$-$5~keV 
flux was (8.38$^{+1.83}_{-1.85}$)$\times$10$^{-15}$ erg cm$^2$ s$^{-1}$. The
reduced $\chi^2$ is 3.1 for 2 degrees of freedom, which corresponds to 
$P(\chi^2)$ = 5\,\%. Although the photon spectrum can formally  
be fitted by a simple power-law function, in the sense 
that $P(\chi^2)$ $\ge$ 5 $\%$, it is apparent that the model 
does not provide the best possible fit to the data;
the observed X-ray spectrum show a concave feature with a break around 
1~keV (Figure 4). Spectral fits with a broken power-law model improve 
the goodness of the fit significantly, as summarized in Table~1. 
Note that, below this break energy, the photon spectrum shows a steep 
power-law index of $\Gamma_{\rm L} \sim $3.0, which  
flattens appreciably to $\Gamma_{\rm H} \sim $1.5 above the break. 
The 0.5$-$5~keV flux was estimated to be 
(8.73$^{+1.85}_{-2.02}$)~$\times$$10^{-15}$ erg cm$^{-2}$ s$^{-1}$.

   \begin{table*}
      \caption[]{Best-Fit Spectral Parameters for knot C of 3C~15}
         \label{tab1}
     $$
         \begin{array}{p{0.1\linewidth}cllll}
            \hline
            \noalign{\smallskip}
Model  & {\rm Photon\hspace{2mm}Index1}^{\mathrm{a}} & {\rm Photon\hspace{2mm}Index2}^{\mathrm{b}}  & {\rm Break E (keV)^{\mathrm{c}}} & {\rm 0.5-5\hspace{2mm} keV\hspace{1mm}flux^{\mathrm{d}}} & {\rm red.\hspace{1mm}\chi^2\hspace{1mm} (dof)}\\
            \hline
PL & 1.71^{+0.44}_{-0.36}  & ...  & ... & (8.38^{+1.83}_{-1.85})\times10^{-15}& 3.1 (2)\\
Broken PL & 3.0 (fixed) & 1.5 (fixed)  & 0.90 \pm 0.18 & (8.73_{-2.02}^{+1.85})\times10^{-15} & 0.98 (2)\\
            \noalign{\smallskip}
            \hline

         \end{array}
     $$
\begin{list}{}{}
\item[$^{\mathrm{a}}$] The best-fit power-law photon index assuming a
                       Galactic $N_{\rm H}$ of 3.03$\times$10$^{20}$ cm$^{-2}$.
\item[$^{\mathrm{b}}$] The best-fit power-law photon index above the break 
                       for a broken power-law model.
\item[$^{\mathrm{c}}$] The best-fit break energy for a broken power-law model.
\item[$^{\mathrm{d}}$] In units of erg cm$^{-2}$ s$^{-1}$ (absorption corrected).

\end{list}
   \end{table*}

   \begin{figure}
   \centering
   \includegraphics[width=7.2cm,angle=90]{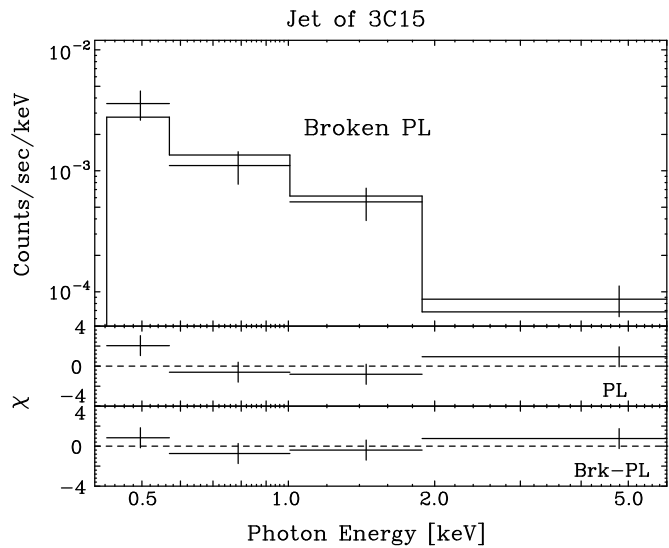}
      \caption{Background-subtracted ACIS spectrum of knot C 
fitted by a broken power-law function. The middle panel shows the residuals to 
the power-law fit with a differential photon index of $\Gamma = $1.7. 
A concave feature is apparent in the residuals. The bottom panel shows the
residuals to the best-fit broken power-law model fixed at  $\Gamma_1$ =
3.0 and $\Gamma_2$ = 1.5.}
         \label{SED}
   \end{figure}

\subsection{X-ray Spectrum of the Lobes}

To examine the diffuse X-ray emission from the radio lobes, 
we accumulated an ACIS spectrum
within an elliptical region containing the whole radio structure.  
An ellipse of 60$''\times$30$''$ centered 
on the nucleus was selected, excluding regions around the jet and the
nucleus (see above) and those within 2$''$ of a 
possible point source. We accumulated the background spectrum 
over a circular region of 1~arcmin radius centered on the nucleus, 
excluding the region of the lobes. Figure~4 shows the background-subtracted ACIS 
spectrum of the lobes of 3C~15 in the 0.5$-$5~keV range. The X-ray
spectrum could not be fitted will by either 
a simple Raymond-Smith (RS) model or a 
power-law (PL) function. The $\chi^2$ were 2.1 and 1.8 
for 6 degrees of freedom, respectively ($P(\chi^2)$ = 0.05, 0.09), 
with a significant excess around 0.7$-$1.0 keV.  
We thus fitted the data by a single PL plus RS model 
with the abundance fixed at 0.4~solar. The obtained parameters are 
summarized in Table 2, and results of fits by PL+RS model is shown in 
Figure 5. The total X-ray flux of the radio lobes is calculated to be 
(3.18$^{+0.75}_{-0.61}$)$\times$10$^{-14}$ erg cm$^{-2}$ s$^{-1}$ in the
0.5$-$5~keV energy range. The X-ray luminosity of the thermal component 
fitted to RS model is  $L_{\rm RS}$ = 
$(4.51^{+2.99}_{-1.91})$$\times$10$^{40}$ erg s$^{-1}$, 
whereas $L_{\rm PL}$ = 
$(1.25^{+0.42}_{-0.46})$$\times$10$^{41}$ erg s$^{-1}$ 
for the non-thermal PL component.

   \begin{table*}
      \caption[]{Best-Fit Spectral Parameters for the Lobes of 3C~15}
         \label{tab2}
     $$
         \begin{array}{p{0.1\linewidth}clllll}
            \hline
            \noalign{\smallskip}
Model  & {\rm kT (keV)}^{\mathrm{a}} & {\rm Abundance^{\mathrm{b}}}  & {\rm RS\hspace{1mm}flux\hspace{2mm}(0.5-5\hspace{2mm} keV)^{\mathrm{c}}}  & {\rm Photon\hspace{2mm}Index}^{\mathrm{d}}  & {\rm PL\hspace{1mm}flux\hspace{2mm}(0.5-5\hspace{2mm} keV)^{\mathrm{e}}} & {\rm red.\hspace{1mm}\chi^2\hspace{1mm} (dof)}\\
            \hline
RS & 2.9^{+1.6}_{-0.9} & 0.4 (fixed) & (2.68\pm0.23)\times10^{-14} & ...  & ... & 2.1 (6)\\
PL & ... & ... &  ... &1.85^{+0.17}_{-0.16}  & (3.07^{+0.35}_{-0.34})\times10^{-14}& 1.8 (6)\\
RS+PL & 0.85^{+0.14}_{-0.08}  & 0.4 (fixed) & (8.70^{+5.80}_{-3.73})\times10^{-15}  & 1.33^{+0.35}_{-0.43} & (2.31 \pm 0.48)\times10^{-14} & 0.41 (4)\\
            \noalign{\smallskip}
            \hline

         \end{array}
     $$
\begin{list}{}{}
\item[$^{\mathrm{a}}$] The best-fit plasma temperature for the Raymond-Smith 
                       model, assuming a Galactic $N_{\rm H}$ of 
                       3.03$\times$10$^{20}$ cm$^{-2}$.
\item[$^{\mathrm{b}}$] Metal abundance for the Raymond-Smith model.
\item[$^{\mathrm{c}}$] Flux of the Raymond-Smith model 
                       in units of erg cm$^{-2}$ s$^{-1}$ (absorption corrected).
\item[$^{\mathrm{d}}$] The best-fit power-law photon index.
\item[$^{\mathrm{e}}$] Flux of the power-law model 
                       in units of erg cm$^{-2}$ s$^{-1}$ (absorption corrected).

\end{list}
   \end{table*}

   \begin{figure}
   \centering
   \includegraphics[width=7.2cm,angle=90]{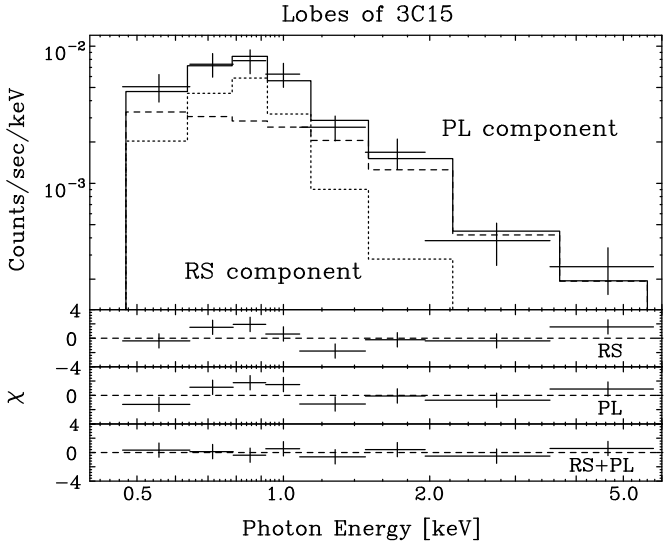}
      \caption{Background-subtracted ACIS spectrum of the lobes of
    3C~15. The dotted and dashed histograms represent the best-fit soft
    RS and hard PL components, respectively. The solid line shows their
    sum. The second panel shows the residuals to the Raymond-Smith fit with a 
temperature of $kT$ = 2.9 keV. The third panel shows the residuals to the 
power-law fit with differential photon index $\Gamma = $1.9. 
The bottom panel shows the residuals to the best-fit Raymond-Smith 
plus power-law model.}
         \label{SED}
   \end{figure}

\subsection{X-ray Spectrum of the Nucleus}

X-ray photons from the nucleus were extracted from a circular 
region of 1.9$''$ radius (within which approximately 98\% of counts 
from a point source would be collected).
Contamination of photons from knot~A$'$ is less than 
1\,\%. The background photons were accumulated over the lobe region. 
Figure~6 shows the background-subtracted ACIS spectrum of the 
nucleus. A high energy bump is seen in the spectrum, 
which cannot be fitted by a simple power-law function 
(red. $\chi^2$ = 32.0 for 22 degrees of freedom; $P(\chi^2)$ = 0.08). 
We therefore adopt a model consisting of two power-law functions, 
one of which is modified by Galactic absorption (PL1) with the 
other (PL2) having a heavily absorbed neutral hydrogen column 
density intrinsic to the source.  A satisfactory fit was obtained if 
we assumed $N_{\rm H}$ = ($7.3_{-3.9}^{+6.0}$)$\times$10$^{22}$ cm$^{-2}$ 
for PL2. In order to reduce the error in the power-law index for PL2, 
we thus fixed $N_{\rm H}$ to 7.0$\times$10$^{22}$ cm$^{-2}$ and refit
the data. The best-fit parameters are summarized in Table~3. 
The absorption corrected fluxes were 
$(6.81_{-1.00}^{+1.14})$$\times$10$^{-14}$ erg cm$^{-2}$ s$^{-1}$ for
PL1, and 
$(4.47_{-1.61}^{+2.40}) \times 10^{-13}$ erg cm$^{-2}$ s$^{-1}$ for 
PL2, respectively. The corresponding X-ray luminosities are 
 $L_{\rm PL1}$ = $(3.52^{+0.59}_{-0.52})$$\times$10$^{41}$ erg s$^{-1}$,  and 
 $L_{\rm PL2}$ = $(2.31^{+1.24}_{-0.83})$$\times$10$^{42}$ erg s$^{-1}$, 
respectively.
 
   \begin{table*}
      \caption[]{Best-Fit Spectral Parameters for the Nucleus of 3C~15}
         \label{tab1}
     $$
         \begin{array}{p{0.1\linewidth}clllll}
            \hline
            \noalign{\smallskip}
Model  & {\rm Photon\hspace{2mm}Index1}^{\mathrm{a}} & {\rm 0.5-5\hspace{2mm}keV\hspace{1mm}flux1^{\mathrm{b}}} & {\rm Absorbed\hspace{2mm}N_H^{\mathrm{c}}} &  {\rm Photon\hspace{2mm}Index2}^{\mathrm{d}}  & {\rm 0.5-5\hspace{2mm} keV\hspace{1mm}flux2^{\mathrm{e}}} & {\rm red.\hspace{1mm}\chi^2\hspace{1mm} (dof)}\\
            \hline
PL & 0.45 \pm 0.09 & (1.37 \pm 0.07)\times10^{-13} & ...  &  .... & ... & 1.45 (22)\\
Dbl-PL & 1.28^{+0.18}_{-0.17} &  (6.81_{-1.00}^{+1.14})\times10^{-14} & 7.0 \times 10^{22}(fixed) & 1.92_{-0.46}^{+0.41} & (4.47_{-1.61}^{+2.40})\times10^{-13}& 0.55 (20)\\
            \noalign{\smallskip}
            \hline

         \end{array}
     $$
\begin{list}{}{}
\item[$^{\mathrm{a}}$] The best-fit power-law photon index for PL1.
\item[$^{\mathrm{b}}$] 0.5$-$5 keV flux for PL1 in units of 
                       erg cm$^{-2}$ s$^{-1}$ (absorption corrected). 
\item[$^{\mathrm{c}}$] Absorbed column density for PL2. 
\item[$^{\mathrm{d}}$] The best fit power-law photon index for PL2.
\item[$^{\mathrm{e}}$] 0.5$-$5 keV flux for PL2 in units of 
                       erg cm$^{-2}$ s$^{-1}$ (absorption corrected).
\end{list}
   \end{table*}

   \begin{figure}
   \centering
   \includegraphics[width=7.2cm,angle=90]{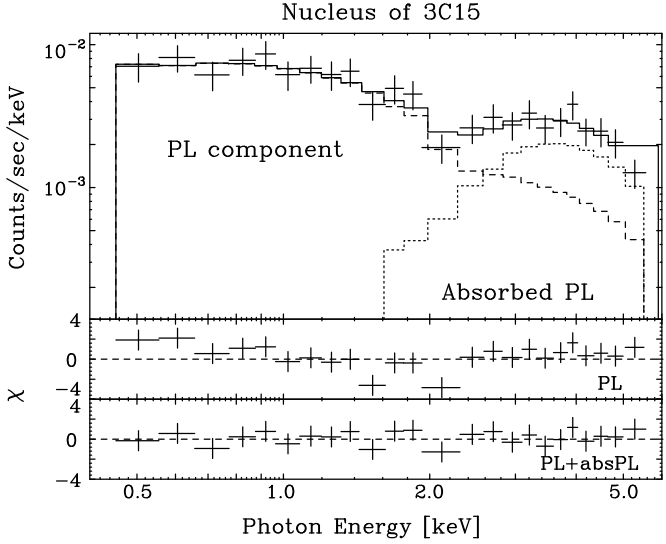}
      \caption{Background-subtracted ACIS spectrum of the nucleus of
    3C~15. The best-fit model, consisting of two power-law components, is
    shown with the histograms. The middle panel shows the residuals to 
the power-law fit with a differential photon index $\Gamma = $0.5. 
The bottom panel shows the residuals to the best-fit double power-law model.}
         \label{SED}
   \end{figure}


\section{Discussion}
\subsection{Emission mechanism of knot C}

As described in the previous sections, we have detected the X-ray 
counterpart of the radio-optical jet-knot in the radio galaxy 3C~15. 
Figure~7 shows the SED from radio to X-ray energies of knot~C. 
Since only the integrated fluxes of the radio and optical knots (from A to 
C or D) are given in the literature, we calculated the radio and optical 
fluxes of knot~C by direct integration from the radio and optical images.
These are 35\,mJy for 8.3\,GHz and 6.5\,$\mu$Jy for $V$ band, respectively  
(see Leahy et al.\ 1997 and Martel et al.\ 1998 for the integrated jet 
fluxes).
Figure~7 clearly indicates that the X-ray flux obtained with $Chandra$ 
is well below the extrapolation from the radio-to-optical continuum. 
A power-law fit ($S_{\nu}$ $\propto$ $\nu^{-\alpha}$, where $S_{\nu}$ 
is the flux density) through the radio and optical points yields a 
spectral index of $\alpha_{\rm ro}$ = 0.9, whereas a 
power-law fit through the optical and X-ray (at 1~keV) data yields a 
steeper spectral index of $\alpha_{\rm ox}$ $\simeq$ 1.1$-$1.2.  
This fact suggests that the synchrotron peak in the
$\nu$$F_{\nu}$ plot must lie below the X-ray energy bands.
However, the X-ray emission process of knot C is currently uncertain due
to the poor photon statistics (Figure~4): in fact, we cannot conclude if 
the X-ray spectrum is ``falling'' or ``rising'' in the $\nu$$F_{\nu}$ 
SED plane. The former case would suggest that the X-rays are produced 
by pure synchrotron emission, as for the radio-to-optical bands, whereas the 
latter would imply that the X-rays are the low end of the hard inverse 
Compton (IC) emission. In the following, we thus consider both these scenarios 
in discussing the jet parameters.

A power-law energy injection of electrons  up to a certain maximum 
energy $\gamma_{\rm max}$ (as expected from shock acceleration) into the 
radiating region should yield a steady-state electron distribution 
approximated by 
\begin{equation}
N(\gamma) = N_0\gamma^{\rm -s}(1 + \frac{\gamma}{\gamma_{\rm brk}})^{-1}
{\rm exp}(- \frac{\gamma}{\gamma_{\rm max}}),
\end{equation}
where $s$ is the injection index, and $N_0$ is the number density of
electrons per energy at $\gamma$ = 1 (e.g., Inoue \& Takahara 1996). 
We assume the minimum Lorentz factor of electrons, $\gamma_{\rm min}$, 
is equal to unity unless otherwise stated (see below).
The parameter $\gamma_{\rm brk}$ is the characteristic energy above 
which the electron spectrum breaks by one power in the index, 
and is determined by the balance between radiative cooling and 
advective escape and/or adiabatic energy loss,
\begin{equation}
\gamma_{\rm brk} = \frac{3 m_{\rm e}c^2}{4(u_B + u_{\rm soft}) \sigma_{\rm T}R} \beta_{\rm sh},
\end{equation}
where $u_B$ is  the magnetic energy density, $u_{\rm soft}$ is the seed 
photon energy density, $R$ is the source radius and 
$\beta_{\rm sh}$ is the shock velocity normalized to the velocity of light. 

The parameter $\gamma_{\rm max}$ could be determined by equating the 
radiative cooling time with the acceleration time scale.  Although the 
acceleration time is not as well understood as the cooling time 
(via the synchrotron and inverse-Compton losses), it can be approximated
 by considering the mean free path, $\lambda(\gamma)$, for the
 scattering of electrons with magnetic disturbances. 
Taking the mean free path to be proportional to the Larmor radius,  
$R_{\rm g}$, by introducing another parameter, $\xi$ 
[=$\lambda$/$R_{\rm g}$], the maximum energy of electrons is given by 
\begin{equation}
\gamma_{\rm max} = [\frac{9 e B}{80(u_B + u_{\rm soft}) \sigma_{\rm T}\xi}]^{1/2} \beta_{\rm sh}.
\end{equation}

We note that the break in the electron population, of course, will not 
appear if $\gamma_{\rm brk}$ is larger than $\gamma_{\rm max}$. In this
case, equation (1) will be reduced to a simpler form 
\begin{equation}
N(\gamma) = N_0\gamma^{\rm -s}{\rm exp}(- \frac{\gamma}{\gamma_{\rm max}}).
\end{equation}
Following Kino \& Takahara 2003, we call the latter situation ``weak cooling
 (WCL)'' since radiative cooling is not efficient, and the 
injected electron spectrum keeps its original form. Meanwhile, 
``moderate cooling (MCL)'' represents the former situation 
($\gamma_{\rm brk}$ $<$ $\gamma_{\rm max}$),  producing a cooling
 break in the electron spectral distribution (equation (1)).

   \begin{figure*}
   \centering
   \includegraphics[width=6.5cm,angle=90]{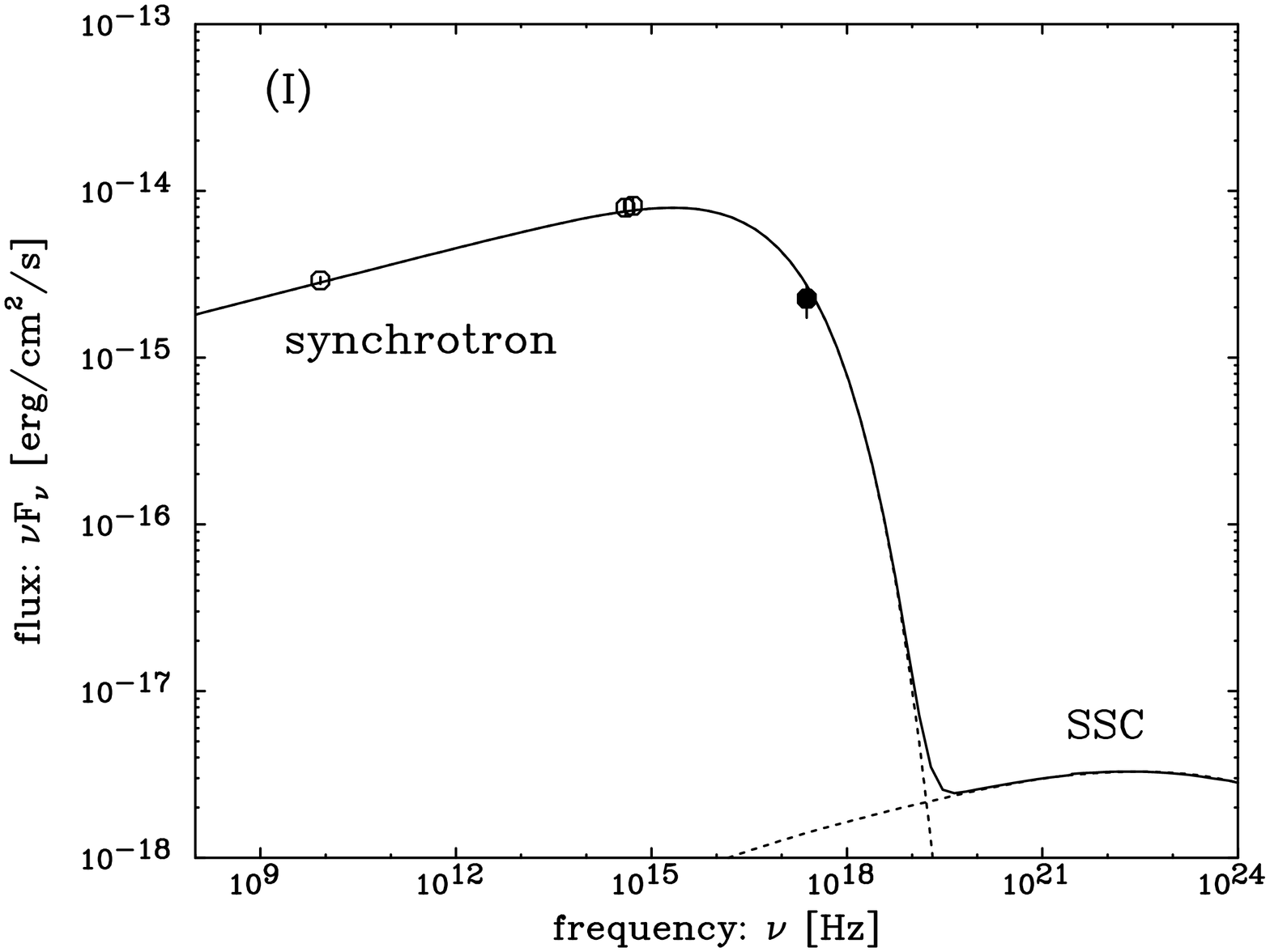}
   \includegraphics[width=6.5cm,angle=90]{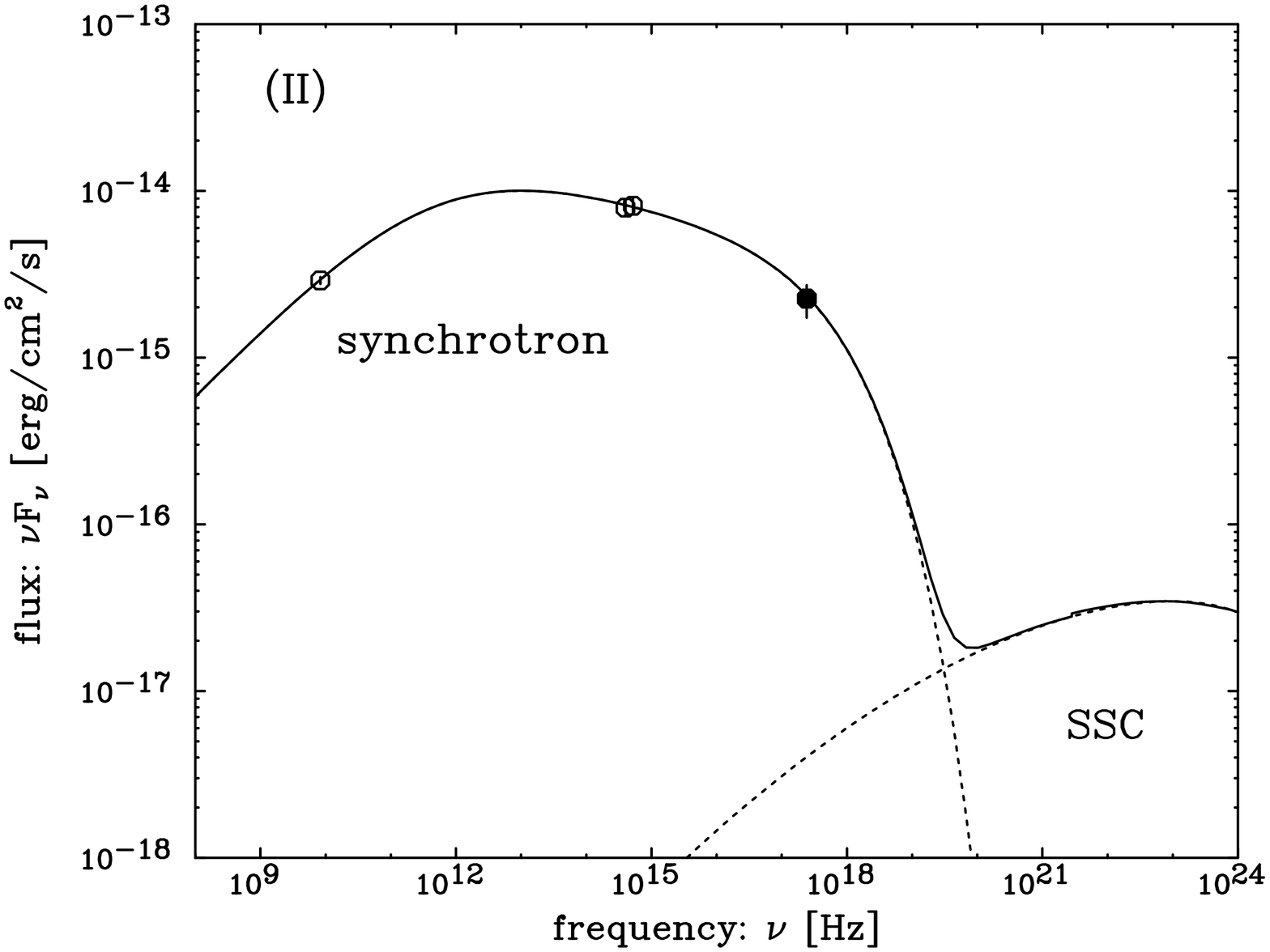}
   \includegraphics[width=6.5cm,angle=90]{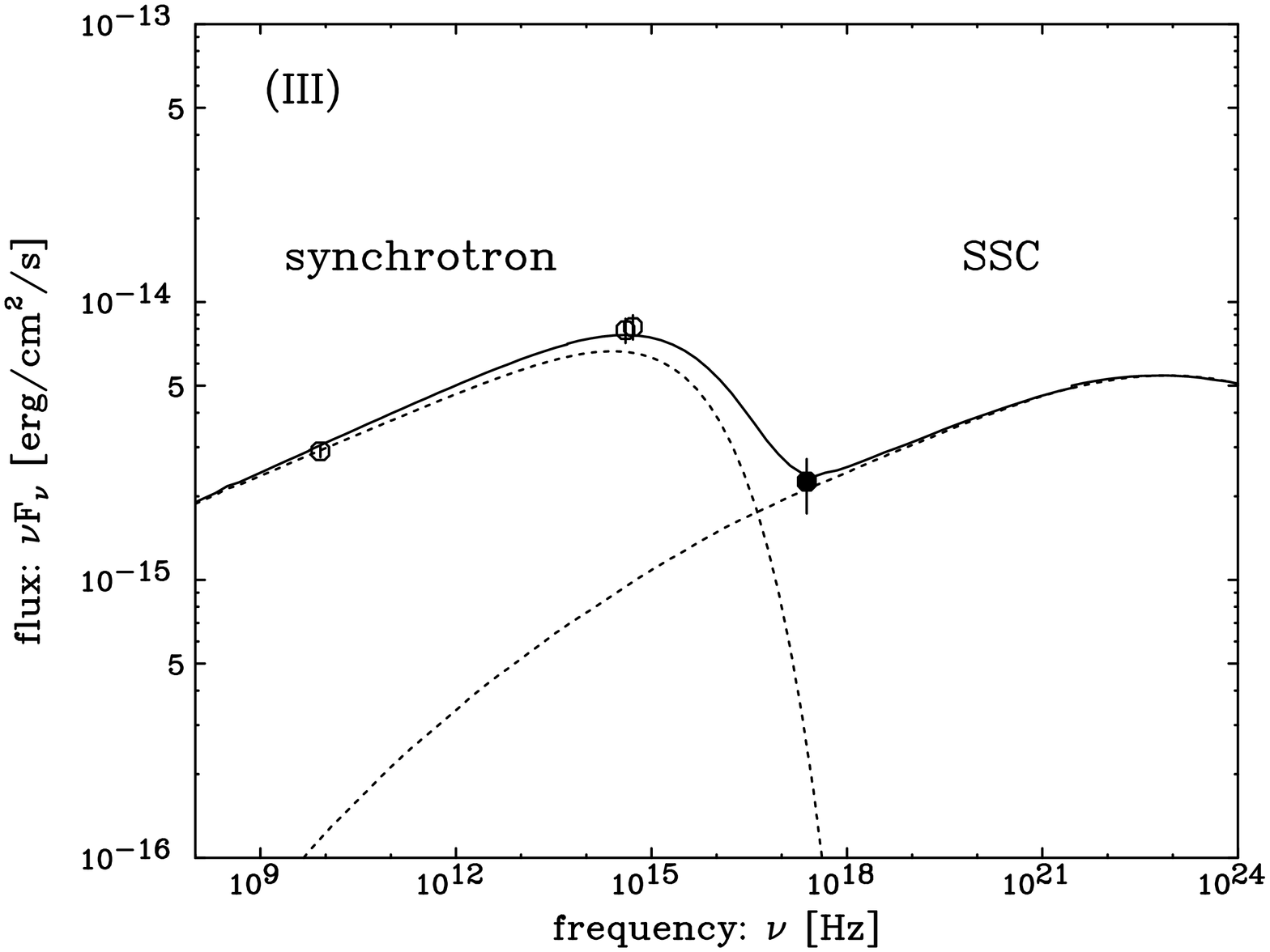}
   \includegraphics[width=6.5cm,angle=90]{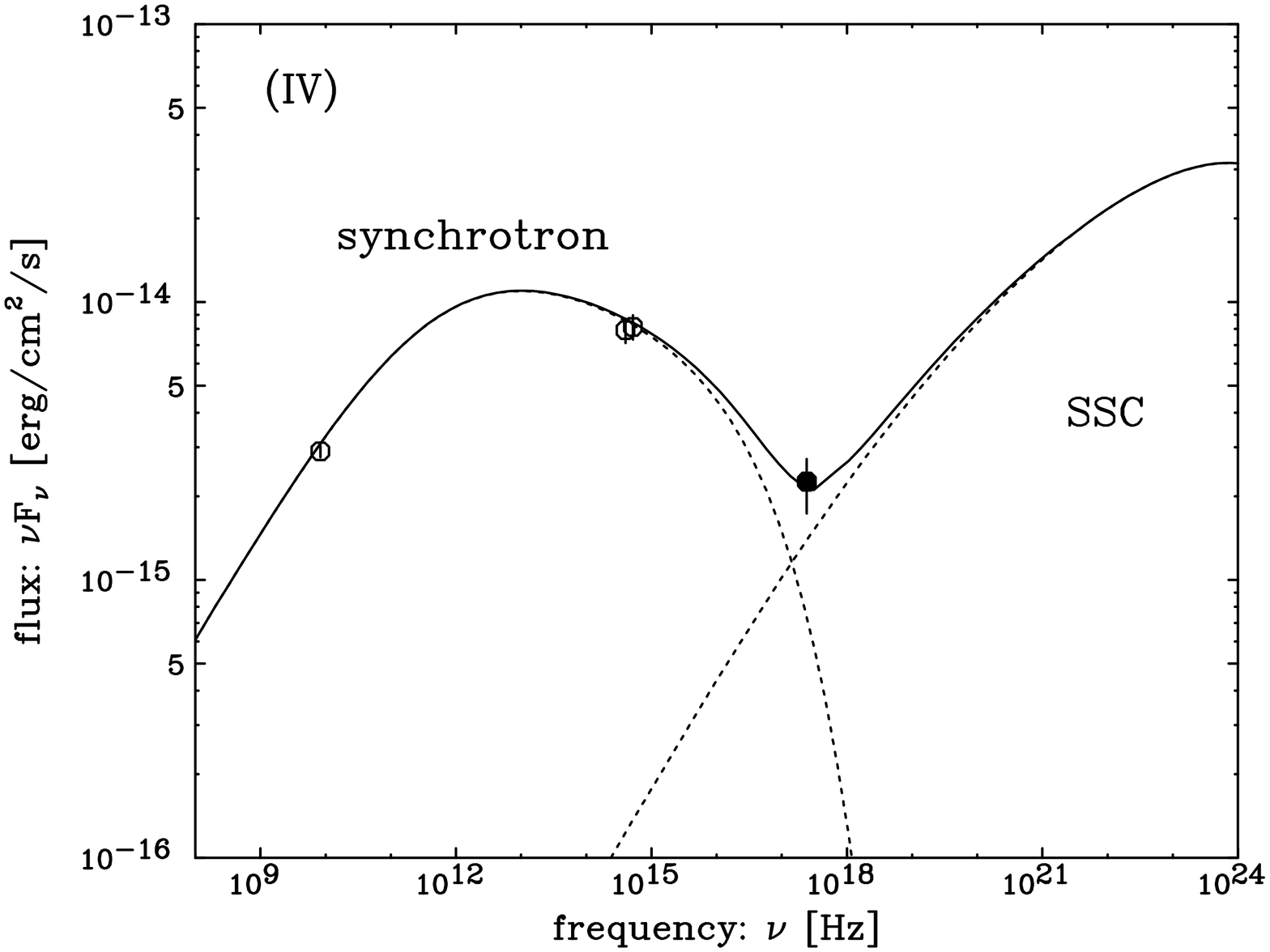}
      \caption{SED of knot C of the 3C~15 jet. The radio and optical fluxes are
    calculated by direct integration from the radio and optical images 
    given in Leahy et al.\ (1997) and Martel et al.\ (1998). 
    The X-ray data are from the $Chandra$ observations
    (this work; 1~$\sigma$ error at 1 keV flux).
    The four models considered in the text are plotted:
    (I) Weak synchrotron cooling for  $u_{\rm e}$ = $u_B$, 
    (II) Moderate synchrotron cooling for $u_{\rm e}$ = $u_B$, 
    (III) Weak synchrotron plus SSC cooling, and 
    (IV) Moderate synchrotron plus SSC cooling. 
    The fitting parameters are summarized in Table~4.}
         \label{SED}
   \end{figure*}

   \begin{table*}
      \caption[]{Comparison of fitting parameters of the SED for knot C}
         \label{tab1}
     $$
         \begin{array}{p{0.2\linewidth}llll}
           \hline
            \noalign{\smallskip}
Parameters  & {\rm (I) Sy (WCL)^{\mathrm{a}}}\hspace{3mm} & {\rm (II) Sy (MCL)^{\mathrm{b}}} & {\rm (III) Sy+SSC (WCL)^{\mathrm{c}}} & {\rm (IV) Sy+SSC (MCL)^{\mathrm{d}}}\\
	  \hline
$R$ (pc)  & 250 & 250 & 250 & 250 \\
$B$ ($\mu$G)  & 570 & 240 & 12 & 8 \\
$N_0$ (cm$^{-3}$)$^{\mathrm{c}}$ & 1.3 \times 10^{-2} & 5.8 \times 10^{-4} & 20.0  & 1.4 \times 10^{-1} \\
$s^{\mathrm{e}}$ & 2.8 & 2.2 & 2.8 & 2.2 \\
$\gamma_{\rm min}$ & 1 & 1 & 1 & 1 \\
$\gamma_{\rm brk}$ & $---$ & 2.6 \times 10^4 & $---$ & 1.5 \times 10^5 \\
$\gamma_{\rm max}$ & 6.0 \times 10^6 & 2.4 \times 10^7 & 1.4 \times 10^7 & 4.0 \times 10^7 \\
$u_{\rm e}$/$u_B$  & 1.0 & 1.0 & 3.6 \times 10^{6} & 2.3 \times 10^{5} \\
$u_{\rm sync}$ (erg cm$^{-3}$) & 1.2 \times 10^{-11} &  1.4 \times 10^{-11}  & 9.7 \times 10^{-12} &  1.4 \times 10^{-11}\\
$L_{\rm kin}$ (erg s$^{-1}$)$^{\mathrm{e}}$ & 1.5 \times 10^{45} \Gamma_{\rm BLK}^2 & 2.6 \times 10^{44} \Gamma_{\rm BLK}^2 & 1.1 \times 10^{48} \Gamma_{\rm BLK}^2 & 3.2 \times 10^{46} \Gamma_{\rm BLK}^2\\
$L_{\rm E}$ (erg s$^{-1}$)$^{\mathrm{f}}$ & 1.9 \times 10^{45} \Gamma_{\rm BLK}^2 & 3.4 \times 10^{44} \Gamma_{\rm BLK}^2 & 1.5 \times 10^{48} \Gamma_{\rm BLK}^2 & 4.3 \times 10^{46} \Gamma_{\rm BLK}^2\\
$\beta_{\rm sh}^{\mathrm{g}}$ & $---$ & 5 \times 10^{-2}  & $---$ & 2 \times 10^{-3}\\
$\xi^{\mathrm{h}}$ &   9.9 \times 10^4 \beta_{\rm sh}^2 & 36.8 & 3.2 \times 10^5 \beta_{\rm sh}^2 & 0.1 \\
$t_{\rm cool}$ (sec)$^{\mathrm{i}}$ & 3.9 \times 10^8 \gamma_{\rm max} \gamma^{-1}\hspace{5mm} &  5.6 \times 10^{8} \gamma_{\rm max} \gamma^{-1}\hspace{5mm} & 1.4 \times 10^{11} \gamma_{\rm max} \gamma^{-1}  &  4.8 \times 10^{10} \gamma_{\rm max} \gamma^{-1}   \\
$c t_{\rm cool}$ (pc)$^{\mathrm{j}}$ & 3.8 \gamma_{\rm max} \gamma^{-1} & 5.4 \gamma_{\rm max} \gamma^{-1} & 1400 \gamma_{\rm max} \gamma^{-1} & 465 \gamma_{\rm max} \gamma^{-1} \\
\hline
CL-brk ? & \times  & \bigcirc & \bigcirc  & \bigcirc \\
$\xi$  $\ge$ 1? & \bigcirc  & \bigcirc & \bigcirc  & \times \\
            \noalign{\smallskip}
            \hline

         \end{array}
     $$
\begin{list}{}{}
\item[$^{\mathrm{a}}$] Corresponding figure is Figure 7 (I).
\item[$^{\mathrm{b}}$] Corresponding figure is Figure 7 (II).
\item[$^{\mathrm{c}}$] Corresponding figure is Figure 7 (III).
\item[$^{\mathrm{d}}$] Corresponding figure is Figure 7 (IV).
\item[$^{\mathrm{e}}$] Kinetic power of electrons in 
		       knot~C. $\Gamma_{\rm BLK}$ is the bulk Lorentz
		       factor of knot~C. 
\item[$^{\mathrm{f}}$] Total power of electrons in knot~C. 
\item[$^{\mathrm{g}}$] Shock velocity of knot~C. 
\item[$^{\mathrm{h}}$] The acceleration parameter given in the
		       text, where $\beta_{\rm sh}$
		       is the shock velocity in the jet. 
\item[$^{\mathrm{i}}$] Cooling time of electrons of energy
		       $\gamma$$m_{\rm e}$$c^2$. 
\item[$^{\mathrm{j}}$] The distance travelled by an electron before losing its
		       energy by radiation.

\end{list}
   \end{table*}

Our goal is to derive the above jet parameters so as not to conflict 
with the observational results. Assuming a spherical geometry for 
the emission region, the radius of the knot C in 3C~15 can be set to 
$R$=7.7$\times$10$^{20}$~cm (corresponding to the FWHM of 0.5~kpc; see 
$\S$2.2). We do not consider Doppler beaming/de-beaming effects
and take $\delta \simeq $1, following the suggestion of Leahy et al.\ (1997).
The injection index of electrons can be determined either by the 
radio-optical index $\alpha_{\rm ro}$ or  optical-X-ray index 
$\alpha_{\rm ox}$: $s$ = 1 $+$ 2$\alpha_{\rm ro}$ = 2.8 for WCL,  
whereas $s$ = 2$\alpha_{\rm ox}$ = 2.2 for MCL. 
We can estimate the synchrotron luminosity integrated over all
frequencies using the formula given by Band \& Grindlay (1985), 
with the result that $L_{\rm sync}$ $\simeq$ 10$^{42}$ erg s$^{-1}$. 
Thus the synchrotron photon energy density is $u_{\rm sync}$ 
$\simeq$ 10$^{-11}$ erg cm$^{-3}$ (see also Table~4). 
Comparing this with the cosmic 
microwave background (CMB) photon energy density boosted in the jet, 
$u_{\rm CMB}$ = 4.1$\times$10$^{-13}$ (1+$z$)$^4$$\Gamma_{\rm BLK}^2$ 
= 5.4$\times$10$^{-13}$$\Gamma_{\rm BLK}^2$ erg cm$^{-3}$, 
the synchrotron photon energy density is more than a factor of
three larger. Here the $\Gamma_{\rm BLK}$ is bulk Lorentz factor of the 
jet and we assume $\Gamma_{\rm BLK}$ $\simeq$ 2.3 
(corresponding to $\beta$ = 0.9 and $\theta$ = $45^{\circ}$; see $\S$ 2.1). 
This indicates that the dominant source of seed 
photons which are upscattered is synchrotron photons, 
$u_{\rm soft}$ $\simeq$ $u_{\rm sync}$ (synchrotron self-Compton (SSC) 
dominated).

In the following, we consider four possible cases to account 
for overall SEDs:\\
(I) Synchrotron X-ray emission under the weak cooling [Sy, WCL]\\
(II) as for (I), but with moderate cooling [Sy, MCL]\\
(III) A composite synchrotron plus SSC X-ray emission under the weak 
cooling [Sync+SSC, WCL]\\
(IV) as for (III), but with moderate cooling [Sync+SSC, MCL]\\

\underline{\bf {(I) Synchrotoron, weak cooling}}\\

We first consider a scenario in which the X-ray emission is purely due to 
synchrotron emission from a power-law electron distribution (equation
(4)). In this case, we cannot determine physical quantities 
uniquely, since the observed SED can be reproduced by any choice
of magnetic field $u_{\rm B}$ or electron energy density $u_{\rm e}$
such that $L_{\rm sync}$ $\propto$ $u_{\rm e}$ $u_B$. 
Here we assume the conventional 
equipartition condition between particles and fields ($u_{\rm e}$ =
$u_B$) for simplicity.  Figure 7 (I) shows the SED fit assuming the 
parameters listed in Table 4. We found that a magnetic field of 
$B_{\rm eq} = $570~$\mu$G is required to reproduce the spectra in 
equipartition, which corresponds to $u_{\rm e}$ = $u_B$ = 
1.3 $\times$ 10$^{-8}$ erg cm$^{-3}$. Comparing this with the
synchrotron photon energy density, $u_{\rm sync}$ = 1.2 $\times$
$10^{-11}$ erg cm$^{-3}$, both the field and particle energy densities
are about 1000 times larger than the radiation energy density.

Since the peak frequency of synchrotron emission of an electron 
of energy $\gamma$$m_{\rm e}$$c^2$ is given by $\nu$ $\simeq$ 10$^6$ 
$B$$\gamma^2$$(1+z)^{-1}$, $\gamma_{\rm max}$ $\simeq$ 6.0 $\times$
10$^6$ is needed to produce X-ray photons. The cooling time of the
highest energy electrons is 3.9$\times$10$^8$~sec, which corresponds to a 
travel distance of $c$$t_{\rm cool} = $3.8~pc, which is much 
$smaller$ than the region size of $R$ = 250\,pc. In such a situation, 
high energy electrons should lose their energy by radiation,  
and hence produce a break in the power-law index at $\gamma_{\rm brk}$ $\sim$ 
0.01 $\gamma_{\rm max}$$\beta_{\rm sh}$. This is inconsistent with our 
original assumption of ``weak cooling'', $\gamma_{\rm brk}$ $\ge$ 
$\gamma_{\rm max}$. There are two different ways of solving this problem. 
We may introduce a cooling break, as is discussed in 
detail in (II), or alternatively, 
the equipartition condition ($u_{\rm e}$ = $u_B$)
may not be met in this source.  The latter assumption requires smaller 
values of $B$, since the highest energy electrons must have a longer life
time. For example, in order for  $\gamma_{\rm brk}$ $\ge$ 10$^{7}$, 
the magnetic field must be $B$ $\le$ 55 $\mu$G (see equation (2)), 
which is an order of magnitude smaller than the equipartition value. 
This latter situation is discussed in detail in (III).\\

\underline{\bf {(II) Synchrotron, moderate cooling}}\\

As we have discussed in (I), equipartition between the electron 
energy density 
and magnetic field 
energy density 
inevitably requires a cooling break since the 
magnetic field is so strong. Here we assume that both the optical and 
X-ray photons are produced by cooled part of the electron population 
($\propto$$\gamma^{-s-1}$). By setting $s$ = 2.2 for the electron 
spectral index, we obtain $\alpha_{\rm R}$ $\simeq$ 0.6 and 
$\alpha_{\rm ox}$ = 1.1, respectively. 
Note that the $\alpha_{\rm R}$ falls in the typical spectral index 
of FR-I/FR-II jets in GHz band 
(0.5$-$0.7; e.g., Bridle \& Perley 1984). 
Extrapolation of these two lines locates the
break frequency at  $\nu_{\rm brk}$ $\sim$ 10$^{12}$ Hz 
$\simeq$ 10$^6$ $B$ $\gamma_{\rm brk}^2$. Under these conditions, 
physical parameters are determined to reproduce the SED (Figure 7 (II)).
A cooling break causes two important effects on the estimation of 
jet quantities;  
(i) substantially reducing electron energy/number density below 
$\gamma_{\rm brk}$, and hence (ii) reducing the equipartition magnetic field 
considerably ($B_{\rm eq}$ = 240~$\mu$G). We found that all the jet 
parameters can be determined self-consistently if the shock velocity 
is sub-relativistic; 
$\beta_{\rm sh}$ $\simeq$ 0.05. The kinetic power, the total power and the 
pressure of the jet are estimated by the following relations:  
\begin{equation}
L_{\rm kin} \simeq \pi R^2 c \Gamma_{\rm BLK}^2 (u_{\rm e}+ u_B),  
\end{equation}
\begin{equation}
L_{\rm E} \simeq \pi R^2 c \Gamma_{\rm BLK}^2 (u_{\rm e} + u_B + P)\hspace{2mm}{\rm ,where} 
\end{equation}
\begin{equation}
P = \frac{1}{3}(u_{\rm e}+ u_B).
\end{equation}
We obtain $L_{\rm kin}$ $\simeq$ 2.6 $\times$ $10^{44}$ $\Gamma_{\rm
BLK}^2$ erg s$^{-1}$, $L_{\rm E}$ = 3.4 $\times$ $10^{44}$ 
$\Gamma_{\rm BLK}^2$ erg s$^{-1}$, and $P$ = 1.5$\times$10$^{-9}$  
erg cm$^{-3}$, respectively. We will compare these jet powers to the 
energy contained in the radio lobes in $\S$ 4.4.

An interesting prediction from this model is that the spatial extent of 
knot~C would be different at different energy bands. This is because 
the radio photons are emitted by the $non$-cooled part of the 
electron population ($\propto$ $\gamma^{-s}$), whereas both the 
optical and X-ray photons are radiated by cooled electrons 
($\propto$ $\gamma^{-s-1}$). 
If the emission region's size is approximately related by 
$R$ $\simeq$ $c$ $t_{\rm cool}$ for cooled electrons, a simple relation 
can be expected for the region sizes; 
$R_{\rm X}$ $<$ $R_{\rm opt}$ $<$ $R_{\rm rad}$. 
Unfortunately, the image resolutions of radio (0.34$''$), 
optical (0.1$''$), and X-rays (0.5$''$) are not sufficient to test this 
hypothesis, but such an approach will provide an interesting 
opportunity for probing the particle acceleration/deceleration mechanism 
in the jet in the near future. \\

\underline{\bf {(III) Synchrotron + SSC, weak cooling}}\\

As we have seen in $\S$3.2, the X-ray spectrum of knot~C shows weak evidence
of mixed power-law (broken power-law) emission. One possibility
to account for the apparent concave feature in the X-ray spectrum 
is the overlap of the synchrotron and SSC components in the
$Chandra$ X-ray energy band. Under this assumption, we first discuss 
``weak cooling'' with an electron spectral index of $s$ = 2.8. 
In principal, we can constrain the physical quantities tightly by 
comparing the synchrotron and SSC luminosity ratio; 
$L_{\rm sync}$ $\propto$ $u_{\rm e}$ $u_B$, 
and $L_{\rm SSC}$ $\propto$ $u_{\rm e}$ $u_{\rm sync}$ .  In order to produce 
$comparable$ synchrotron and inverse Compton luminosities, a large
departure from an equipartition is unavoidable since $u_{\rm B}$ 
$\sim$ $u_{\rm sync}$. Assuming the observed value of 
$u_{\rm sync}$ ($\sim$ 10$^{-11}$erg cm$^{-3}$), we require a magnetic 
field of an order $B$$\sim$ 10 $\mu$G. In order to explain the observed 
synchrotron spectrum with such a weak magnetic field, the electron
energy density $u_{\rm e}$ must be large, since $u_{\rm e}$ $\propto$ $L_{\rm
sync}$/$u_B$. A more accurate calculation provides a satisfactory fit 
to the SED for $B$ $\simeq$ 12 $\mu$G  ($u_B$ = 5.7$\times$10$^{-12}$ erg 
cm$^{-3}$)  and $u_{\rm e}$ = 2.1$\times$10$^{-5}$ erg cm$^{-3}$,  
respectively (Figure 7(III)). 

Note, however, that the SSC luminosity is quite uncertain because we have to 
infer the overall SSC spectra from the ``bottom edge'' of the hard 
X-ray spectrum. 
We thus consider whether or not other choices of magnetic field could
also reproduce the data.  For a given synchrotron luminosity 
($L_{\rm sync}$) in a given region size ($R$), the SSC peak frequency 
and luminosity vary as $\nu_{\rm ssc}$ $\propto$ 
$B^{-1}$ and $L_{\rm ssc}$ $\propto$ $B^{-2}$, respectively. 
Therefore, variations in $B$ shift the position of the SSC peak 
as $L_{\rm ssc}$ $\propto$ $\nu_{\rm ssc}^2$. In order to achieve
comparable synchrotron/SSC fluxes  in the $Chandra$ X-ray band, only a 
magnetic field of $B$ $\simeq$ 10 $\mu$G is acceptable.
In such a situation, the cooling time of the highest energy electrons
($\gamma_{\rm max}$ = 1.4$\times$10$^7$) is 1.4$\times$10$^{11}$~sec, 
which corresponds to a travel distance of 1400~pc. This suggests that 
$\gamma_{\rm brk}$ does $not$ appear as long as $\beta_{\rm sh} \ge
$0.2. In contrast to the case (I), this is consistent with our 
original assumption of weak cooling.
 
While this model (III) reproduces the observed concave feature of 
X-ray spectrum, there are two major difficulties in understanding 
the derived jet quantities.  
First, the departure from equipartition, $u_{\rm
e}$/$u_B$ $\simeq$ 3.6$\times$10$^6$, seems to be too large. One 
possibility to reduce the discrepancy is to adopt a larger value of 
$\gamma_{\rm min}$ ($\gg$~1), since the electron energy density behaves 
$\propto \gamma_{\rm min}^{-s+2}$. For example, if we assume 
$\gamma_{\rm min}$ $\simeq$ 10$^3$, then $u_{\rm e}$/$u_B$ would be 
$\sim$ 1.4$\times$10$^4$. This is still far from equipartition, though the 
difference is significantly reduced. (Note, however, that such a large value 
of $\gamma_{\rm min}$ is unlikely from the viewpoint of shock 
dynamics in jets, where $\gamma_{\rm min}$ $\sim$ $\Gamma_{\rm BLK}$; 
Kino \& Takahara 2003). Therefore, a large departure from equipartition 
cannot be avoided within the framework of  model (III).  
The second problem is related with unreasonably large jet power expected 
with this model. The total power of the jet is given by $L_{\rm E}$ = 
1.5 $\times$ $10^{48}$ $\Gamma_{\rm BLK}^2$ erg s$^{-1}$. This 
jet power exceeds the total output of most bright quasars (e.g., Celotti,
Padovani, \& Ghisellini 1997), though 3C~15 is a relatively weak radio 
source (intermediate between FR-I and FR-II; Leahy et al.\ 1997).

Also if the jet X-ray emission were of inverse Compton origin, we would 
expect it to follow the synchrotron emission more closely (emission from
low-energy electrons not affected by radiative loss), whereas in fact 
it is closer to the optical, suggesting that X-ray emissions are 
dominated by synchrotron emission. Therefore, we consider model (III) 
to be less plausible than (II), given the very weak evidence from 
the X-ray spectrum. In $\S$ 4.4, we will discuss this model from the
view point of the energetic link between the jet and lobes.
Future observations at the radio-IR bands ($10^{11-13}$~Hz), as well as 
higher quality X-ray data are strongly encouraged to test the 
predictions
of model (III). \\

\underline{\bf {(IV) Synchtotron + SSC, moderate cooling}}\\

Finally we consider the case in which the electron distribution has 
a cooling break and most of the X-ray photons are produced via SSC. 
Following case (II), we assume $s$ = 2.2 and the break frequency of 
$\nu_{\rm brk}$ $\sim$ 10$^{12}$ Hz. Figure 7 (IV) shows the SED fit using 
this model. The existence of a cooling break substantially reduces 
the electron energy density, $u_{\rm e}$, and hence reduces the discrepancy 
between $u_{\rm e}$ and $u_B$. However, the magnetic field strength is still 
30 times below the equipartition value of $B_{\rm eq}$ = 240 $\mu$G.
The total power of the jet is estimated to be $L_{\rm E}$ = 4.3 $\times$
$10^{46}$ $\Gamma_{\rm BLK}^2$ erg s$^{-1}$, which is reduced by a factor 
of 40 from model (III). In such a situation, the shock velocity must be
non-relativistic; $\beta_{\rm sh}$ $\simeq$ 2$\times$10$^{-3}$.

Although this model (IV) seems to reproduce the overall SED well, 
it requires an acceleration parameter of smaller than unity 
($\xi$ = 0.1). In the conventional picture of resonant 
pitch angle scattering (Blandford \& Eichler 1987), $\xi$ can be 
identified with the ratio of energy in the non-turbulent magnetic field to 
that in the turbulent field. Thus, $\xi$ is expected to be larger than 1 
by definition. For example, Inoue \& Takahara (1996) argued that 
$\xi$ $\gg$ 1 for sub-pc--scale jets in blazar type AGNs, whereas 
in normal plasmas such as that in the interstellar medium or 
supernova remnants, $\xi$ is inferred to be of order unity 
(e.g., Bamba et al.\ 2003). We thus consider $\xi$$<$1 is 
unreasonable in the framework of diffusive 
shock acceleration theory.\\

\underline{\bf {Summary of Model Fitting for Knot C}}\\

In summary, we have considered four possible models to reproduce the
overall SED of knot~C in 3C~15. The  major results of our
considerations are:\\
(i) The synchrotron X-ray model under equipartition inevitably requires a 
    cooling break, since high energy electrons lose their energy on 
    very short time scales ($t_{\rm cool}$ $\le$ $R/c$). \\
(ii) A composite synchrotron plus SSC model predicts a magnetic 
    field far below equipartition ($B_{\rm eq}$ $\gtrsim$ 30 $B$).\\  
Taking these into account, we consider case (II) to be the 
most likely for the overall SED of knot~C (Sy, MCL). 
Although case (III) is still possible, the large departure from 
equipartition, as well as the extremely large jet power, $L_{\rm E}$ $>$ 
10$^{48}$ erg s$^{-1}$, seems to be problematic. In all four models, 
(I)$-$(IV), electrons must be accelerated up to 
$\gamma_{\rm max}$$\sim$ 10$^7$ in knot~C of 3C15, 
which corresponds to a travel distance of 
$ct_{\rm cool}$ $\le$ 1.4\,kpc (see Table~4). 
Comparing this with the $projected$ distance of knot~C from the nucleus, 
5.1~kpc, re-acceleration is clearly necessary in knot~C. 

\subsection{Emission mechanism of the Lobes}

In our 30~ksec $Chandra$ observation of 3C~15, we have detected  
diffuse X-ray emission closely associated with the radio lobes.
The X-ray spectrum can be reproduced by a two component model consisting of 
soft and hard components. Considering the temperature and the luminosity 
of the Raymond-Smith fit, the soft X-ray component may naturally be 
attributed to thermal emission from the hot halo of the host galaxy 
(e.g., Matsushita et al.\ 2000). In fact, the X-ray luminosity of 
thermal component,  $L_{\rm RS}$ = 4.5$\times$10$^{40}$ 
erg s$^{-1}$, is well within the typical range 
of elliptical galaxies, 10$^{39-42}$ erg s$^{-1}$ 
(e.g., Eskridge et al.\ 1995). 

The power-law modeling of the hard component gives a spectral index 
$\alpha_{\rm X}$ = 0.33$^{+0.35}_{-0.43}$. Unfortunately we cannot 
compare it with the synchrotron radio index of 3C~15, since we have 
the radio-lobe flux at only one frequency (i.e.\ the 3.6~cm map). 
However, it is known that the $integrated$ radio spectrum has 
$\alpha_{\rm R}$ = 0.75 from the NED photometry data base 
(http://nedwww.ipac.caltech.edu/). Since the total radio flux is 
almost entirely dominated by the lobes at radio frequencies (e.g., 
76\,\% of the radio flux is due to lobe emission at 3.6~cm), 
it could be a reasonable approximation of synchrotron radio emission
from radio lobes. Figure 8 shows the SED of radio lobes 
thus produced. The NED data, multiplied by a factor 0.76, are nicely
consistent with the lobe flux estimated at 3.6 cm. 
We can clearly see that the radio flux does not connect smoothly 
with the flat X-ray spectrum. We thus consider that the diffuse hard 
X-rays are produced via inverse Compton scattering on the electrons which emit 
the synchrotron photons in the radio band. The slightly different spectral 
index between the radio and X-rays may imply that X-ray photons are
produced by $lower$ energy electrons than those which produce the
observed radio emission.
In fact, Kellermann et al.\ (1969) reported that the spectral index 
does flatten somewhat at lower frequencies.
 
In order to determine the origin of the seed photons which are upscattered 
to X-ray energies, we approximate the geometry of the radio lobes as a 
cylinder 62.5~kpc (50$''$) in length and 20~kpc (16$''$) in diameter. 
This yields a lobe volume of $V$ = 5.8$\times$10$^{68}$~cm$^3$ and 
a synchrotron photon energy density of $u_{\rm sync}$ $\sim$ 
3.5$\times$10$^{-14}$ erg cm$^{-3}$ (see Table 5). 
Similarly, IR/optical photons from the host galaxy halo are expected 
to provide an energy density of 
2$\times$10$^{-13}$ (10 kpc/$d$)$^2$ erg cm$^{-3}$, where $d$ is 
the distance from the nucleus to the X-ray emitting region in the lobe 
(Sandage 1973). Since the energy density of these seed photons falls 
below that of the CMB, $u_{\rm CMB}$  =  5.4$\times$10$^{-13}$ erg 
cm$^{-3}$, we will consider CMB photons as the dominant seed photons 
for inverse Compton X-ray emission. Note, however, that IR/optical 
emission may dominate over the CMB in the innermost regions of the radio
lobe, $d \le $5~kpc. In this case, it may be that inverse 
Compton scattering of different seed photons is occurring in the 
X-ray (CMB seed photons) and $\gamma$-ray (IR/optical seed photons) bands.

   \begin{figure}
   \centering
   \includegraphics[width=6.5cm,angle=90]{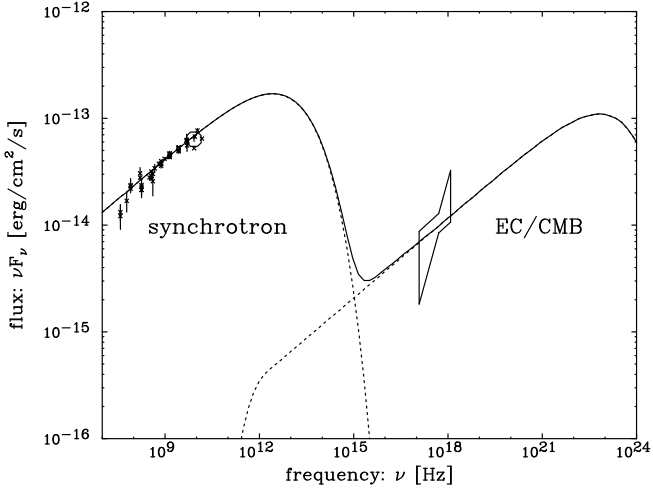}
      \caption{SED of radio lobes in 3C~15. $Bow$ $tie$: X-ray power-law
    spectrum obtained with $Chandra$ (this work), 
    $Open$ $circle$: obtained by direct integration
    from the 3.6 cm radio image in Leahy et al. (1997). 
    $Crosses$: NED data base, but multiplied by 0.76.}
         \label{SED}
   \end{figure}

   \begin{table*}
      \caption[]{Fitting parameters of the SED for radio lobes}
         \label{tab1}
     $$
         \begin{array}{p{0.4\linewidth}l}
           \hline
            \noalign{\smallskip}
Parameters  & {\rm Sy + EC/CMB}\\
	  \hline
$V$ (cm$^3$)$^{\mathrm{a}}$  & 5.8 \times10^{68} \\
$B$ ($\mu$G)  & 3.7\\
$N_0$ (cm$^{-3}$) & 7.0 \times 10^{-4}\\
$s$ & 2.5\\
$\gamma_{\rm min}$ & 1\\
$\gamma_{\rm max}$ & 1.0 \times 10^6\\
$u_{\rm e}$/$u_B$  & 2.1 \times 10^{3}\\
$u_{\rm sync}$ (erg cm$^{-3}$) & 3.5 \times 10^{-14}\\
$P_{\rm lobe, non-th}$\hspace{5mm}(erg cm$^{-3}$)$^{\mathrm{b}}$ & 3.8
 \times 10^{-10}\\
$E_{\rm lobe}^{\mathrm{c}}$ (ergs)  & 8.9 \times 10^{59}\\
$t_{\rm cool}$ (sec) & 5.3 \times 10^{13} \gamma_{\rm max} \gamma^{-1}\\
$c t_{\rm cool}$ (pc) & 5.2\times 10^5 \gamma_{\rm max} \gamma^{-1}\\
            \noalign{\smallskip}
            \hline
         \end{array}
     $$
\begin{list}{}{}
\item[$^{\mathrm{a}}$] Volume of the radio lobes.
\item[$^{\mathrm{b}}$] Pressure from non-thermal electrons in the lobes.
\item[$^{\mathrm{c}}$] Total energy supplied in the radio lobes.
\end{list}
   \end{table*}

The ratio of the radio (synchrotron) flux to the X-ray 
(inverse Compton) flux is therefore related by
\begin{equation}
\frac{f_{\rm X}}{f_{\rm radio}} \simeq \frac{u_{\rm CMB}}{u_B}.
\end{equation}
Here, we must compare the radio/X-ray fluxes which correspond to 
the $same$ population of relativistic electrons. The Lorentz factors 
of electrons which upscatter CMB photons ($\nu_{\rm CMB}$$\simeq$ 
10$^{11}$\,Hz) to the X-ray band ($\nu_{\rm X}$ $\simeq$
10$^{17-18}$\,Hz) would be $\gamma$ $\simeq$10$^{2.9-3.4}$. 
These electrons emit synchrotron photons in the frequency range 
of $\nu$ $\simeq$ 10$^{6-7}$ ($B$/1$\mu$G)\,Hz, for which there is no 
observational data (Figure 8).  We thus extrapolate the observed radio 
flux by assuming the radio spectral index of $\alpha_{\rm R} \simeq  
$0.75. We obtain $f_{\rm radio}$ = 2.1 $\times$ 10$^{-14}$
($B$/1$\mu$G)$^{1-\alpha_{\rm R}}$ erg cm$^{-2}$ s$^{-1}$ for the 
frequency range of $\nu$ $\simeq$ 10$^{6-7}$ ($B$/1$\mu$G)\,Hz. 
Comparing this with the integrated X-ray flux of $f_{\rm X}$ = 
(2.3$\pm$0.5)$\times$ 10$^{-14}$ erg cm$^{-2}$ s$^{-1}$ in the 
$Chandra$ bandpass, the magnetic field strength is estimated to be 
$B$ $\simeq$ 4\,$\mu$G. The result of SED model fitting is shown as solid
lines in Figure 8. We have assumed the maximum Lorentz factor of electrons 
to be $\gamma_{\rm max}$ $\simeq$ 10$^6$. 
The best fit parameters are summarized in Table 5. 

A comparison of the synchrotron and inverse Compton flux densities 
allows us to derive $u_{\rm e}$ = 1.1$\times$10$^{-9}$ erg cm$^{-3}$ 
$\simeq$ 2100 $u_B$.  Thus we find that there is a significant electron 
dominance in the lobes of 3C~15. This ratio is larger than those
reported for lobes in other radio galaxies ($u_{\rm e}$/$u_B$ $\le$ 100; 
e.g., Tashiro et al.\ 1998; Isobe  2002), but the difference is 
mainly due to the different choice of $\gamma_{\rm min}$ : we have assumed 
$\gamma_{\rm min}$ = 1 throughout this paper as discussed in $\S$4.1.
If we set $\gamma_{\rm min}$ $\simeq$ 1000 as for these published works, 
we obtain $u_{\rm e}$ $\simeq$ 66 $u_B$, which is well in the range of 
published works. 
The total power supplied in the lobes is estimated as:
\begin{equation}
E_{\rm lobe} = (u_{\rm e} + u_B)V  + PV \simeq
8.9 \times 10^{59} {\rm ergs}. 
\end{equation}

It is interesting to compare the thermal pressure estimated 
from the Raymond-Smith (RS) X-ray emission, and non-thermal pressures 
of relativistic electrons in the radio lobe. 
If we assume a very simplified model in which 
thermal photons are produced by electrons which have a constant number 
density, $n_{\rm e, th}$ over the radio lobes, 
the best fit X-ray spectral model 
(see $\S$\,3.3 and Table~2) provides the number density of thermal
electrons as 
\begin{equation} 
n_{\rm e, th} = 3.1 \times 10^{-3}\hspace{5mm}{\rm cm^{-3}}, 
\end{equation} 
where we have assumed that the thermal plasma volume is equal to that of 
radio lobes; $V$ = 5.8$\times$$10^{68}$ cm$^3$. The pressure of thermal 
plasma is thus given by
\begin{equation} 
P_{\rm lobe, th} = 1.9 n_{\rm e, th} kT = 8.1 \times 10^{-12}\hspace{5mm}{\rm erg} \hspace{1mm} {\rm cm^{-3}}, 
\end{equation} 
where $kT$ = 0.85 keV from  $\S$ 3.3 (see Table 2).
Meanwhile, the non-thermal pressure of relativistic electrons is estimated as 
\begin{equation} 
P_{\rm lobe, non-th} = \frac{1}{3}(u_{\rm e} + u_B) = 3.8\times10^{-10}\hspace{5mm}{\rm erg} \hspace{1mm} {\rm cm^{-3}}. 
\end{equation} 
The large difference between $P_{\rm lobe, non-th}$ and $P_{\rm lobe, th}$
may partly due to our over-simplified model for the thermal plasma
distribution, but it seems that a 
more powerful (dense) thermal plasma, other than that 
emitting the X-rays, is necessary to confine the relativistic electrons in 
the radio lobes.

Finally, we should comment on the features of the surface brightness ratio 
($S_{\rm radio}$/$S_X$) given as Figure~2(c). It has been claimed that 
this ratio represents the magnetic field distribution along the major 
axis of the lobes, since $S_{\rm radio}$/$S_{\rm X}$ $\propto$ $B^{2}$ 
(except for the nucleus regions; see equation (8)). 
In this sense, the magnetic field may be weakest near the nucleus, 
increasing several-fold at larger distances, 
as was suggested for the field-particle distribution of the FR~II radio 
galaxy 3C~452 (Isobe et al.\ 2002). However, we should note that the 
thermal component, if 
distributed over the scale of the galaxy, could be centrally condensed. 
These are possibly related to the thermal RS emission seen 
in the X-ray spectrum of the radio lobes (Figure 5). In principle, 
it would be possible to discriminate between the two emission components by 
comparing X-ray images in different energy bands, where the soft X-rays 
($\le$ 2 keV) are dominated by thermal RS emission and the hard X-rays 
($\ge$ 2 keV) would be of non-thermal inverse Compton origin. 
Unfortunately, photon statistics are not sufficient for such a study. 
We simply comment here that the apparent center filled morphology of 
the X-ray image 
(Figure 2 (a)) may partly be due to a centrally condensed thermal plasma, 
rather than variations in the magnetic field strength. The actual
inverse-Compton profile of the radio lobes might be closer to the synchrotron 
radio profile than suggested in Figure 2. Future deep X-ray observations
will clarify this point further.
  
\subsection{Emission mechanism of the Nucleus}
In $\S$3.4, we showed that the X-ray spectrum of the nucleus consists
of two different power-law components. The low-energy component is well 
represented by a hard power-law with Galactic
absorption, whereas the high-energy component is heavily absorbed with 
$N_{\rm H}$ = 10$^{22-23}$~cm$^{-2}$.  One possibility to account for 
this X-ray spectrum is that the two components are produced in 
different emission regions around the nucleus: 
the low-energy power-law may be due to the inner-jet emission and
possibly comes from the parsec-scale region, 
whereas the high-energy component may be
emission from the hidden active nucleus at the center.   
Considering the symmetrical lobe morphology, the obscuring 
material is most likely the molecular torus fueling the nucleus which is
postulated to exist in active galaxies (e.g., Antonucci \& Miller 1985; 
Krolik \& Begelman 1986).

Interestingly, additional evidence of a hidden active nucleus has been 
obtained from the surface brightness profile of 3C~15 in the $V$ and $R$ 
bands. While these are very well described by a de Vaucouleurs profile, 
$\propto$ $r^{1/4}$,  between 0.5~kpc and 10~kpc, 3C~15 lacks a sharp, 
point-like AGN inside a radius of 0.3$''$ (Martel et al.\ 1998). 
This optical observation is consistent with the lack of strong 
AGN-type features in the spectrum of 3C~15 (Tadhunter et al.\ 1993). 
Two possibilities have been suggested to account for the absence of a 
resolved AGN, which is in contrast common to the other 3C radio galaxies 
with optical jets imaged with $HST$. First, the central source may be 
presently inactive, but is activated at regular intervals, leading to 
the periodic ejection of material along the jet axis 
(resulting in knots A, B, and C).  
Alternatively, heavy obscuration by dust towards the line of sight 
to the nucleus could also account for the nondetection of an AGN. 
Dust permeates the central regions of 3C~15: if the extinction is 
large enough, the bright nucleus would then be hidden from our line 
of sight as implied from the X-ray spectrum of the nucleus. 
It should be noted that 
the dust seen in the $HST$ images is on a much larger scale than the 
hypothetical molecular torii around AGNs.  Thus it might be possible
that the dust cloud orbiting on the kpc-scale that we directly 
see is the material which obscures part of the X-ray core; in other
words, a large ``dust torus'' might be present around the nucleus of 
3C~15.

\subsection{The link between the jet and lobes}

Comparison of the power transported by the jet ($L_{\rm E}$) and 
the total energy contained in the lobes ($E_{\rm lobe}$) provides an 
interesting opportunity to probe the dynamics in the jet. By dividing 
$E_{\rm lobe}$ by $L_{\rm E}$, we can estimate the ``fueling time'' of
the jet $t_{\rm fuel}$, which can be compared with the source age
$t_{\rm src}$ (see $\S$1). In model (II) of $\S$4.1, we obtained 
$L_{\rm E}$ $\simeq$ 3.4 $\times$ $10^{44}$ $\Gamma_{\rm BLK}^2$ 
erg s$^{-1}$.  The fueling time is expected to be $t_{\rm fuel}$ = 
8.2$\times$$10^{7}$ $\Gamma_{\rm BLK}^{-2}$~yrs. 
This is qualitatively consistent with the source age of radio lobes, 
$t_{\rm src}$ $\sim$ 2.1$\times$$10^7$(0.01 $c$/$v_{\rm exp}$)~yrs. 
In contrast, other models, in particular (III), predict  
significantly smaller $t_{\rm fuel}$ compared to $t_{\rm src}$.
In fact,  for $L_{\rm E}$ $\simeq$ 1.5 $\times$ $10^{48}$ $\Gamma_{\rm BLK}^2$ 
erg s$^{-1}$, we obtain $t_{\rm fuel}$  = 1.9 $\times$ 10$^4$ yrs.

These facts suggest again that the synchrotron emission with moderate 
cooling, case (II), provides a reasonable explanation of 
the X-ray emission mechanism of knot~C.
However, there are still several possibilities to account for the
apparent discrepancy in model (III), i.e., $t_{\rm fuel}$ $\ll$ $t_{\rm src}$. 
First, we may have $underestimated$ the total energy in the lobes since 
we only take account of relativistic electrons. In fact, we cannot 
estimate the contribution from thermal electrons and possibly protons 
which should constitute a reservoir for acceleration of non-thermal 
electrons. 
If only 0.1\% of electrons 
are ``visible'' in the radio and X-ray bands, the discrepancy may be 
reduced significantly. (However, we note that the situation is the 
same for the total power of the jet, $L_{\rm E}$, in the sense that we only 
take account of the contribution from ``visible'' relativistic electrons.)

Alternatively, it is plausible that by considering the brightest jet 
knot (knot~C) we have $overestimated$ the jet power. The knots are 
thought to be where the particle and/or field densities are 
enhanced compared to the rest of the jet. It may be reasonable 
to assume that the jet is actually very sparse, and that its average 
power is much lower, e.g., 0.1\% of that in the bright knot. It might 
be said that the jet has a spatial filling factor of only $\sim$10$^{-3}$. 
Such uneven jet structure may be related with the activity of the 
central nucleus. If the nucleus expels blobs of material 
intermittently, rather than in a stationary manner, it will produce 
bright knots along the jet axis at semi-regular intervals 
(such as that observed for knots A, B, and C).  
This idea can be readily accommodated
by the internal shock scenario, recently proposed 
by a number of authors to account for the variability properties of 
sub-pc scale jets (e.g., Spada et al.\ 2001, Kataoka et al.\ 2001, 
Tanihata et al.\ 2003). 
Although somewhat speculative, occasional activity of the jet 
(0.1\% of the time) may explain the energy link between 
the jets and the lobes in radio galaxies quite well. Future observations in 
the hard X-ray and gamma-ray bands are necessary to test this further.

\section{Conclusion}
We have reported the X-ray detection of the jet, lobes, 
and absorbed nucleus of the FR~II radio galaxy 3C~15. 
The X-ray image obtained with $Chandra$ clearly shows that most 
of the X-ray jet emission comes from knot~C, which was previously detected
in the radio and optical jets. We found that the X-ray flux is well below 
the extrapolation from the radio-to-optical continuum. We consider four
possible cases to reproduce the overall SED: 
(I) Weak synchrotron cooling with $u_{\rm e}$ = $u_B$, 
(II) Moderate synchrotron cooling with $u_{\rm e}$ = $u_B$, 
(III) Weak synchrotron plus SSC cooling, and 
(IV) Moderate synchrotron plus SSC cooling. 
Case (I) and (IV) were safely ruled out by considering the acceleration
and radiative cooling processes over the source. We argue that case (II) 
is a reasonable scenario to understand the SED of knot~C, but 
the case (III) is possible if equipartition is strongly violated 
($u_{\rm e}$/$u_B$ $\simeq$ 3.6$\times$10$^6$) and the jet power is 
extremely large ($L_{\rm E}$ $\simeq$ 10$^{48}$ erg s$^{-1}$). 
In either case, the highest energy electrons, with $\gamma_{\rm max}$ 
$\ge$ 10$^7$, need to be re-accelerated in  knot~C. 
The diffuse hard X-ray emission associated with the lobes 
is most likely due to the inverse Compton emission of CMB photons by 
the synchrotron emitting electrons in the radio lobes. Assuming a radio 
spectral index of $\alpha_{\rm R}$ $\simeq$ 0.75, we found that the 
lobes in 3C~15 are particle dominated, where $u_{\rm e}$/$u_B$ $\sim$ 2100. 
The fueling time ($t_{\rm fuel}$) is qualitatively consistent with the 
source age $t_{\rm src}$ for the case (II), whereas the latter must 
be significantly shorter for the case (III). The discrepancy of (III), 
however, can be understood if the jet is actually very sparse, and has a
spatial filling factor of only 10$^{-3}$. 
By comparing  the thermal pressure associated with the galaxy halo and 
non-thermal pressure in the radio lobe, we found that relativistic
electrons cannot be confined only with the X-ray emitting thermal gas. 
Finally, we show that the X-ray emission from the nucleus consists of 
two power-law components, one of which suffers from 
significant absorption $N_{\rm H}$ $\simeq$ 
10$^{22-23}$ cm$^{-2}$ intrinsic to the source. 
Such a high column density may support the existence of dusty 
torus around the nucleus, as suggested by the optical observations. 

\begin{acknowledgements}
We appreciate the insightful comments and suggestions of the referee
, Dr. M. Georganopoulos, that have improved this paper.
We also thank Dr. N. Isobe for useful discussion of X-ray properties of radio
 lobes. J.K.\ acknowledges a Fellowship of the Japan Society
for Promotion of Science for Japanese Young Scientists.
This research has made use of the NASA/IPAC Extragalactic Database (NED)
 which is operated by the Jet Propulsion Laboratory, California
 Institute of Technology, under contract with the National Aeronautics 
and Space Administration.
\end{acknowledgements}

\end{document}